\documentclass{aa}

\usepackage{graphicx}
\usepackage{txfonts}
\usepackage{natbib}
\usepackage{xspace}
\usepackage{color}
\usepackage{hyperref}

\usepackage{threeparttable}
\usepackage{supertabular}
\usepackage{multirow}
\usepackage{makecell}

\usepackage{amsmath}
\usepackage{caption,subcaption}         
                               
\usepackage{upgreek}
\usepackage{float}

\usepackage{lipsum}
\usepackage{tablefootnote}
\usepackage{tikz}
\usepackage{lscape}             
                                
\usepackage{placeins}           
\usepackage{adjustbox}
\newcommand{\cref}{\ref}
\usepackage{orcidlink}

\newcommand{\ie}{\textit{i}.\textit{e}. }

\newcommand{\xspec}{\texttt{Xspec}\xspace}
\newcommand{\bwcycl}{\texttt{bwcycl}\xspace}
\newcommand{\maxi}{MAXI \xspace}
\newcommand{\exo}{EXO~2030+375\xspace}

\newcommand{\rxj}{RX~J0440.9+4431\xspace}
\newcommand{\fuo}{4U~0115+63\xspace}
\newcommand{\herx}{Her~X-1\xspace}
\newcommand{\cenx}{Cen~X-3\xspace}
\newcommand{\lmcx}{LMC~X-4\xspace}
\newcommand{\grojs}{GRO~J1750-27\xspace}

\newcommand{\lumcgs}{erg~s$^{-1}$}
\newcommand{\lumcgsts}{$\times10^{37}$~erg~s$^{-1}$}

\newcommand{\msol}{\ensuremath{M_{\odot}}\xspace}
\newcommand{\rsol}{\ensuremath{R_{\odot}}\xspace}

\newcommand{\mdot}{\ensuremath{{\dot m}}\xspace}

\newcommand{\pcm}{\,$\mathrm{cm^{-2}}$}	
 
\newcommand{\cmsq}{\,$\mathrm{cm^{2}}$}

\newcommand{\kev}{\,$\mathrm{keV\,}$}
\newcommand{\kpc}{\,$\mathrm{kpc\,}$}

\newcommand{\gpcmsqps}{\,$\mathrm{g\,s^{-1}\,cm^{-2}}$}

\newcommand{\Tmound}{T_{\rm th}}

\newcommand{\vmound}{v_{\rm th}}

\newcommand{\taumound}{\tau_{\rm th}}

\newcommand{\taumax}{\tau_{\rm max}}

\newcommand{\tautrap}{\tau_{\rm trap}}

\newcommand{\cpl}{\texttt{cutoffpl}}
\newcommand{\hcut}{\texttt{highecut}}
\newcommand{\tbabs}{\texttt{TBabs}}
\newcommand{\bbodyrad}{\texttt{bbodyrad} }
\newcommand{\bbody}{\texttt{bbody}}
\newcommand{\comptt}{\texttt{compTT}}

\newcommand{\cyclabs}{\texttt{cyclabs}}
\newcommand{\npex}{\texttt{NPEX}}

\def\sig{\sigma_{_{\rm T}}}
\def\sigpar{\sigma_\|}
\def\sigperp{\sigma_\perp}
\def\sigbar{\overline\sigma}

\def\nh {N${\rm_H}$ }
\def\kt {kT${\rm_{BB1}}$ }
\def\ktt {kT${\rm_{BB2}}$ }
\def \hxmt {\emph{Insight-HXMT} }
\def \nustar {\emph{NuStar} }
\def \nicer {\emph{NICER} }
\def\astrosat {\emph{AstroSat} }
\def \fermi {\emph{Fermi} }

\begin{document}

 \title{Insight-HXMT Spectral and Timing Studies of a Giant Outburst in \rxj }
   \titlerunning{The Giant Outburst of \object{\rxj}}

   \author{Prahlad R. Epili \inst{1}\,\orcidlink{0000-0002-2278-8375}
   \and Wei Wang\inst{1}\,\orcidlink{0000-0003-3901-8403}}
   \institute{Department of Astronomy, School of Physics and Technology, 
             Wuhan University, Wuhan 430072, China P.R.C \\
             \email{prahlad@whu.edu.cn, wangwei2017@whu.edu.cn}
           }

   \date{Received September 30, 20XX}

  \abstract
   {The Be/X-ray binary pulsar \rxj underwent a giant outburst in late 2022 and lasted three months. The \hxmt\  has observed this source at several instances of the entire outburst in 2022-2023. We used these bright outburst observations of the pulsar to study its X-ray spectral and timing variability. The pulse profiles obtained at similar luminosity during the progress and declining phases of the outburst show a similar shape behavior. With the increase in source luminosity, the complex pulse profile with multiple peaks at low luminosity becomes a single peaked pulse profile at the high luminosity at the outburst peak. The phase-averaged spectra of the pulsar in 1-120 \kev are explained with an absorbed cutoff power-law continuum model. 
   During the outburst phases, we have found the evidence of a cyclotron resonance scattering feature in the spectra varying in energies ($\sim33.6-41.6$)\kev having broad linewidth > 5\kev. In declining phases of the outburst, we have also found the hints of first cyclotron harmonic varying in 65-75~\kev. The application of thermal and bulk Comptonization model to the phase-averaged and phase-resolved spectra reveals a high surface magnetic field ($B\sim10^{13}$~G) for the pulsar.}


   \maketitle

\section{Introduction}
The Be/X-ray binary \rxj was first discovered during a ROSAT Galactic plane survey along with its optical companion as LS~+44~17/BSD~24-491 \citep{1997A&A...323..853M}. X-ray pulsations from the source with a period of $\sim202.5$~sec were identified by \citet{1999MNRAS.306..100R}. The first evidence of pulsar outburst activity in X-rays was found by \citet{2010ATel.2527....1M} with the \maxi instrument in late March 2010. The peak luminosity of this first outburst in 3-30\kev was $3.9\times10^{36}$\lumcgs \citep{2012PASJ...64...79U}, suggesting the source to be transient in nature with a relatively high quiescent luminosity, $L_{x}\sim10^{34}$\lumcgs. The recent distance estimate to the source as $2.44^{0.06}_{0.08}$~\kpc \citep{2021AJ....161..147B} puts the observed luminosity of the source far below $10^{37}$\lumcgs typical for the Type I outbursts of Be X-ray binaries \citep{2011Ap&SS.332....1R}.
\maxi/GSC detected a brightening of an X-ray source located at \rxj  eleven years after the third outburst \citep{2022ATel15835....1N}. Interestingly, towards the end of this outbreak, its count rate increased again \citep{2023ATel15868....1P}, reaching a peak X-ray flux of 2.25 Crab \citep{2023ATel15907....1C}, as recorded by Swift /BAT (15-50 \kev). This unprecedented event allows for the study of the accretion process of \rxj across a wider range of luminosity with high quality data. 

The soft X-ray spectrum of \rxj as observed with different X-ray observatories in the past, can be explained with a power-law plus blackbody components in addition to a 6.4\kev iron line \citep{2012PASJ...64...79U,2012A&A...539A..82L,2021AJ....161..147B}. In hard X-rays, an absorption feature at $\sim30$\kev is reported as a possible cyclotron line \citep{2012MNRAS.421.2407T}. However, there were speculations for existence of this spectral feature in the later studies by \citet{2013A&A...553A.103F}. The high quality broadband observations are necessary to confirm the presence or absence of cyclotron resonance scattering feature (CRSF) in the source. Given the recent re-brightening of the pulsar, \hxmt observations cover many luminosity epochs of the outburst. These can be studied to better understand the pulsar broadband spectrum with physical and empirical continuum models. In the present work, we have used these high quality broadband observations of the pulsar to study its pulse phase-averaged and phase-resolved spectra. 

In this paper, we briefly present our recent results from timing studies and pulse phase-averaged and phase-resolved spectral studies of the pulsar with an empirical continuum model and physics based Comptonization model. In this work, we analyze the observations of this source by \hxmt\  to characterize its temporal properties, with a focus on the evolution of pulse profile that is energy and luminosity dependent. From the high-energy data that benefited from \hxmt, we can further understand the emission beam pattern of accretion columns in different accretion states.

\section{Observations and Data Reduction}\label{sec:obs}

The \hxmt is China's first space-based X-ray satellite capable of observing X-ray sky in 1--250\kev \citep{2020SCPMA..6349502Z}. The three main X-ray instruments onboard \hxmt are: the high energy X-ray telescope (HE, 20-250\kev pointing with effective area $\sim5100$\cmsq) \citep{2020SCPMA..6349503L}, the medium energy X-ray telescope (ME, 8$\sim$35\kev, 952\cmsq)\citep{2020SCPMA..6349504C} and the low energy X-ray telescope (LE, 1$\sim$12\kev, 384\cmsq)\citep{2020SCPMA..6349505C}.

\begin{figure}
 \begin{center}
 \includegraphics[height=2.4in, width=3.6in, angle=0]{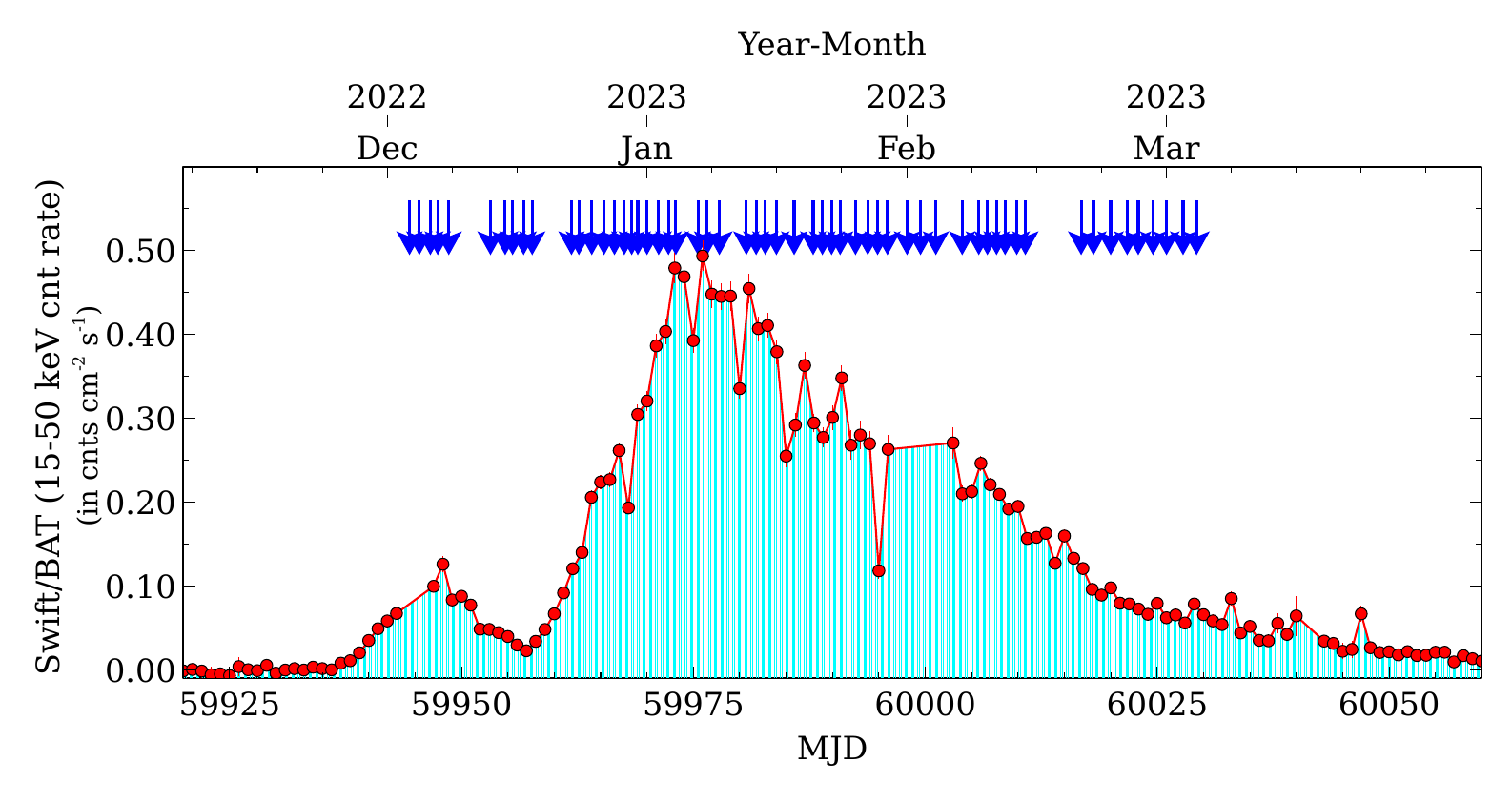} 
 \end{center}
 \vspace{-15pt}
\caption{\small The long term Swift/BAT light curve of \rxj in 15-50\kev covering the 2022-2023 giant outburst duration. The arrows in the insets of the figure represent the start of pointed observations of \rxj performed with \hxmt. The error-bars represent 1$\sigma$ uncertainties. A log of these pointed observations is shown in Appendix~\ref{tab:log1}. }
\vspace{-15pt}
\label{lsv-swift-bat}
\end{figure} 
\rxj has been observed with \hxmt during a giant outburst in late December-2022 to March 2023. \citet{2023MNRAS.526.3637L} has used these outburst observations to study its timing properties of the pulsar. The phase-averaged spectral evolution of the pulsar at three different luminosity epochs has been studied \citep{2024A&A...689A.316L}. However, given the many epochs of the observations by \hxmt during the giant outburst, it is interesting to study its timing properties and spectral evolution using all the \hxmt observations. In Figure~\ref{lsv-swift-bat}, we have shown the \hxmt pointed observations of \rxj across the Swift/BAT 15-50~\kev long-term lightcurve covering the outburst duration. The log of these observations are provided in the Appendix~\ref{tab:log1}.

To analyze the observations, we have used the \hxmt data analysis software package, \texttt{hxmtsoft-v2.05} to generate the cleaned event files. For the calibration of cleaned event data, we have used the \texttt{caldb v2.06} provided by the instrument team. In order to generate the data products such as spectra and light curves of the pulsar in different energies, we have used the following screening criteria to obtain good time intervals (GTIs) with the tasks \emph{hegtigen, megtigen, legtigen} to the cleaned event files:  1) elevation angle > $10^{\circ}$, 2) pointing offset angles < $0.04^{\circ}$, 3) geomagnetic cutoff rigidity > 8 GeV and eliminating the time intervals corresponding to the passage of South Atlantic Anomaly. The screening event tasks \emph{hescreen, mescreen, lescreen} were used to screen the data. The arrival times of all the screened event data were subjected to the Solar System barycenter correction with the task \texttt{hxbary}. We have used the tasks \emph{helcgen, melcgen, lelcgen} with a bin size of 1/128 ~s to extract the source and background light curves in different energy range. The source spectra have been extracted with the tasks \emph{hespecgen, mespecgen, lespecgen} respectively for HE (30-120\kev), ME(10-30\kev) and LE(1-10\kev). The respective background products such as spectra and light curves are estimated using the tasks such as \emph{hespecbkg, mespecbkg, lespecbkg} and \emph{helcbkg, melcbkg, lelcbkg} respectively. For the broadband spectral analysis of \hxmt\ observed spectra of \rxj, we have used \xspec v12.11.1 \citep{1996ASPC..101...17A}. All uncertainties in the spectral parameters are obtained at a confidence level of 68\%.

\section{Results}

\subsection{Timing studies}
In order to estimate the spin period of the pulsar from the \hxmt observed light curves, we have combined ME light curves in 10-30\kev from  all the exposure IDs of a given observation ID.  We have applied the epoch-folding method (\emph{efsesarch} task of \emph{FTOOLS}) to obtain the spin period of the pulsar for each of the observed epochs. We note that, our estimates of the pulsar spin period is within the error estimates of pulsar spin period as obtained by \citet{2023MNRAS.526.3637L}. Therefore we have used these spin period values of the pulsar to construct the pulse profiles in LE (1--10\kev), ME(10--30\kev) and HE (30-150\kev) for each epoch using the epoch-folding FTOOL \emph{efold}. The folding epoch for each observation is adjusted manually near the start of each observation to obtain two complete pulse profiles. The energy resolved pulse profiles obtained during the progress and declining phases of the outburst at similar luminosity are compared and shown in Figure~\ref{lsv-profiles}. It can be seen from the Figure~\ref{lsv-profiles} that, the pulse profiles at the low values of X-ray luminosity are complex and show multiple peaks. However, with the increase in source luminosity, the pulse profiles evolve and become single peaked at the peak of the giant outburst. In hard X-rays (\ie 30--150\kev), this transition of multiple peaked pulse profile at low luminosity to a prominent single peaked pulse profile occurs at luminosity value of $\sim2.5$\lumcgsts.

\subsubsection{Pulse-profiles at similar luminosity}
\begin{figure*}
 \begin{center}$
 \begin{array}{lcr}
 \includegraphics[height=2.8in, width=2.37in, angle=0]{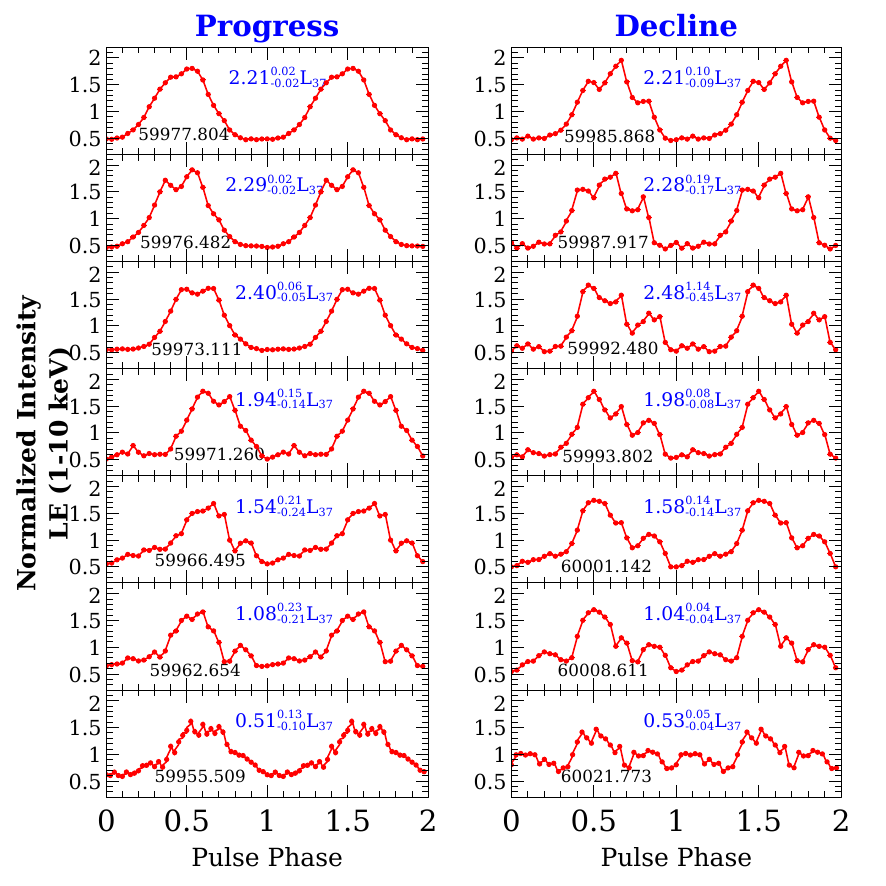} &
 \includegraphics[height=2.8in, width=2.37in, angle=0]{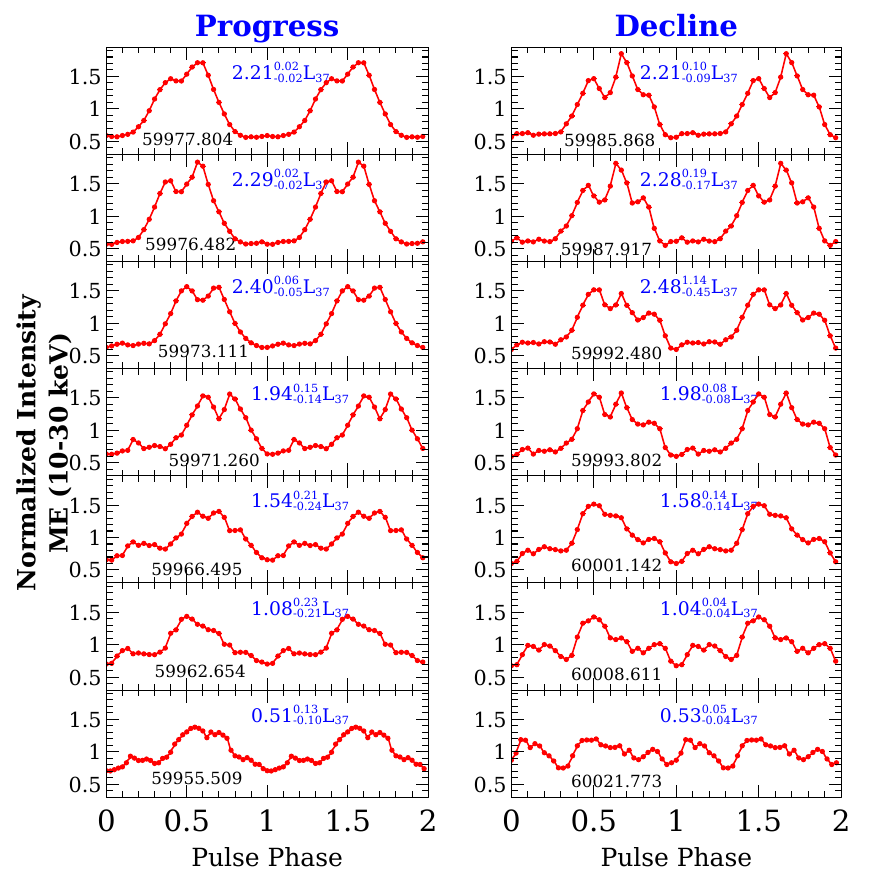} & 
 \includegraphics[height=2.8in, width=2.37in, angle=0]{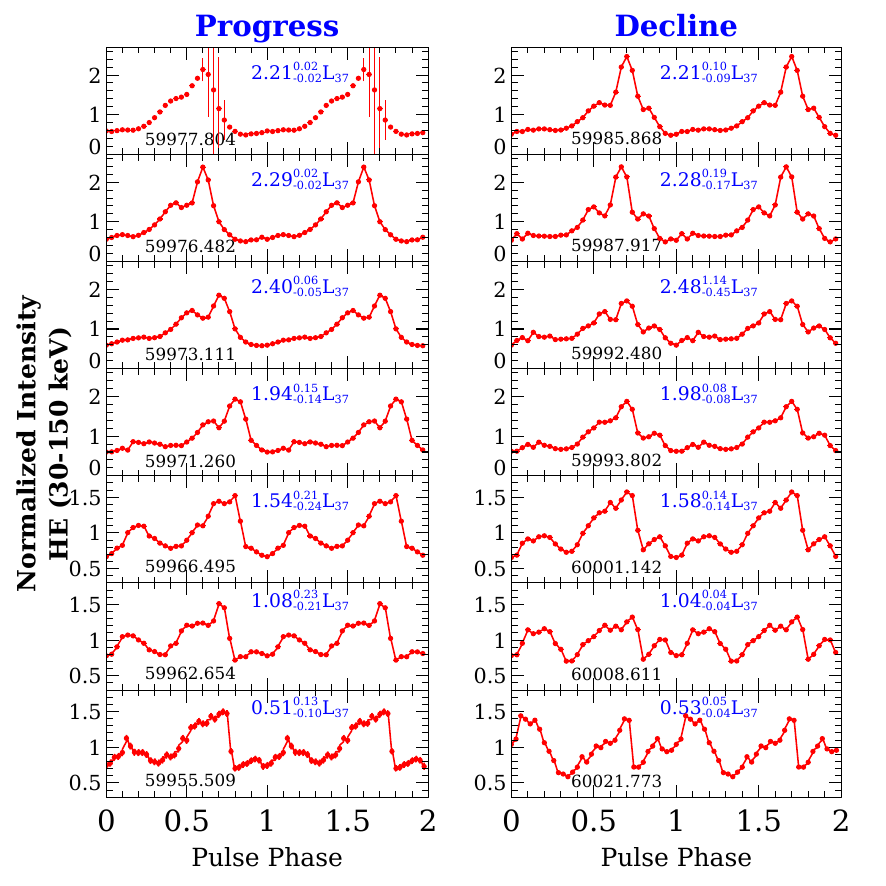} 
 \end{array}$
 \end{center}
 \vspace{-15pt}
\caption{\small Energy resolved pulse profiles of \rxj obtained by folding the light curves from \hxmt LE (1--10\kev) (2 panels from left), ME (10--30\kev) (2 panels in the middle) and HE (30-150\kev) (2 panels from right) light-curves at similar luminosity during the progress and decline of the giant outburst. Two pulses are shown in each panel for clarity. The error-bars represent 1$\sigma$ uncertainties. The epochs of the observations in MJD are shown in inside each panels on the left. The corresponding source luminosity  estimates in 1-120\kev is shown in the middle of each panels.}
\label{lsv-profiles}
\end{figure*}

In order to check the luminosity dependence of the pulse profiles, we have obtained the pulse profiles of the pulsar in 1-10\kev, 10-30\kev \& 30-150\kev at various luminosity epochs.  At similar luminosity during the progress and decline of the outburst these pulse profiles are compared. It is found that, the hard X-ray pulse profiles 30-150\kev are found to be similar in shape at both the declining and inclining phases of the outburst at similar luminosity epochs. These pulse profiles are shown in the Figure.~\ref{lsv-profiles}. However, we observe that the pulse profiles obtained at declining phases of the outburst shows evolution of substructures at the primary peak of the pulse profile. In particular the ME and HE pulse profiles at low luminosity (\ie $\sim0.5$\lumcgsts) show two major peaks respectively near 0.1-0.2 pulse phase and 0.7 pulse phase. With the increase in source luminosity, the first peak gradually diminishes in size, leading to the secondary sub-peak as the prominent peak in the pulse pulse profiles at higher source luminosity.

\subsubsection{Pulse fraction variations}
To measure the changes in the pulsar intensity during the giant outburst at different luminosities, we have computed the pulse fractions across the entire outburst in three different energy ranges (\ie 1--10 \kev, 10--30 \kev and 30--150 \kev). The pulse fraction is defined as : \(PF = (I_{max}-I_{min})/(I_{max}-I_{min})\), where $I_{min}$ and $I_{max}$ are the minimum and maximum intensity in the pulse profile. A variation of the pulse fraction values across the entire outburst is shown in Figure~\ref{lsv-pf}. In the Appendix~\ref{tab:log1}, a complete log of these pulse fraction changes of the pulsar is shown. It can be seen from Figure~\ref{lsv-pf}(a), that the pulse fraction (PF) increases gradually as the source luminosity increases. The PF at low energy (1--10 \kev) increases from $\sim44$\% at the onset of the outburst (\ie at MJD 59944.4) to a maximum value of $\sim64.2$\% near the outburst peak (\ie at MJD 59983.9). Subsequently the low energy PF decreases to $<27\%$ at MJD 60029.25 with the decline of the outburst. Similar trend is seen in the PF values in medium energy ME pulse profiles (\ie in 10--30\kev). The PF in ME lightcurves shows an upward trend from 30.4\%  at the beginning of the outburst (\ie at MJD 59944.4) to a maximum value of 58.7\% at the outburst peak (\ie at MJD 59980.7). Subsequently, with the decline of the outburst, it reaches a minimum PF of 20\% (at MJD 60029.3). The PF changes in hard X-rays as seen from HE lightcurves (\ie in 30-150 \kev), varies from $\sim36\%$ at the start of the giant outburst (near MJD 59955.5) reaching a maximum PF of $71.5\%$ at MJD 59982.8. Thereafter, it gradually decreases to $27.7\%$ at MJD 60010.8 before the outburst fades. We find that, interestingly, the hard X-ray PF further increases during the fading phases of the giant outburst to a maximum of $42\%$ within the \hxmt observations during which the LE and ME PF values show a decreasing trend with the decline of the outburst. 
\begin{figure*}[ht]
    \centering
    \begin{subfigure}{0.37\textwidth}
        \centering
        \includegraphics[width=\textwidth]{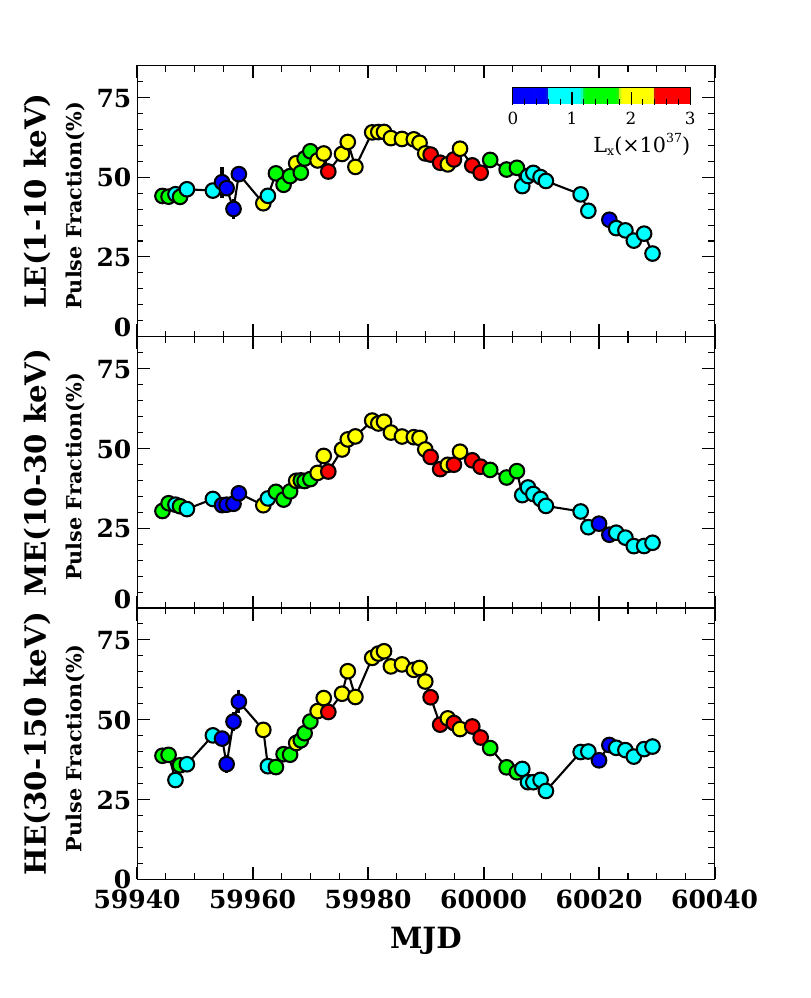}
         \vspace{-20pt}
    \end{subfigure}
    \begin{subfigure}{0.37\textwidth}
        \centering
        \includegraphics[width=\textwidth]{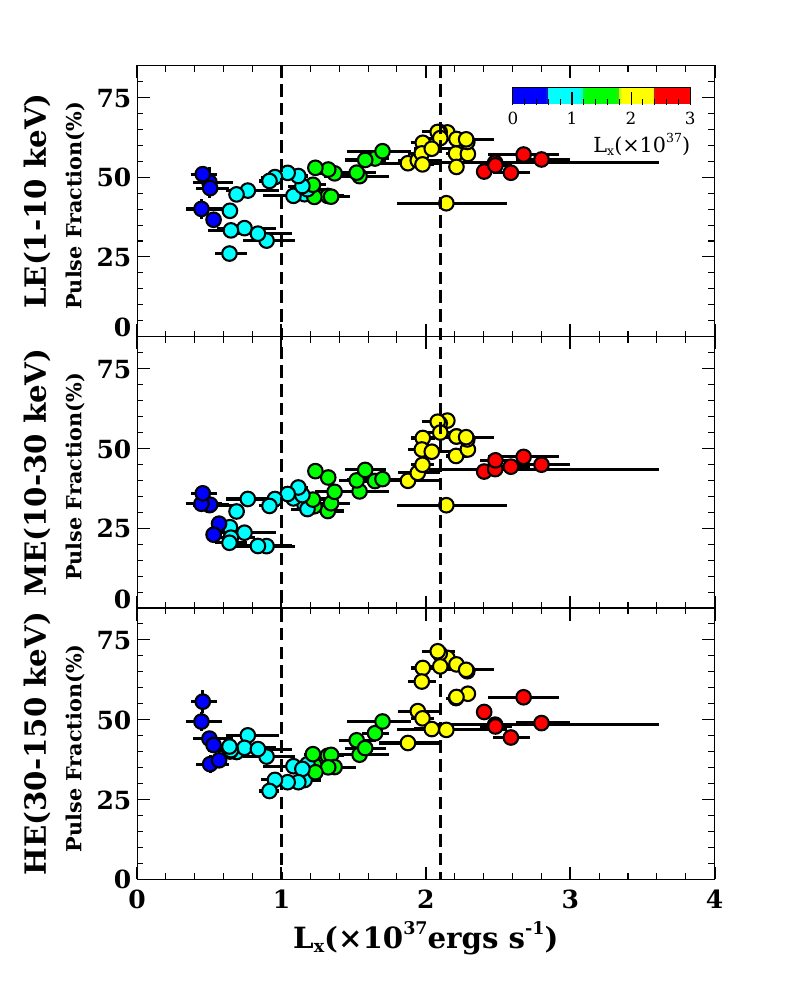}
        \vspace{-20pt}
    \end{subfigure}
\vspace{-5pt}    
\caption{\small {\it Left:} (a) The variation of pulse fractions in \rxj during the 2022-2023 giant outburst in the energy ranges of 1-10\kev, 10-30\kev and 30-150\kev obtained respectively from the LE, ME and HE light curves. 
{\it Right:} (b) The variation of pulse-fraction is shown with the outburst source luminosity in 1--120\kev in \rxj during the 2022-2023 giant outburst in the energy ranges of 1-10\kev, 10-30\kev and 30-150\kev obtained respectively from the LE, ME and HE light curves. Table~\ref{tab:log1} in the Appendix shows values of these pulse fraction estimates and the respective epochs of the giant outburst. The color bar represents the 1--120\kev source luminosity in the units of $10^{37}$\lumcgs. The two vertical dashed lines at $L_{x}\sim1$\lumcgsts and $L_{x}\sim2.2$\lumcgsts represent the turn-over points in the variation of PF with outburst luminosity}    
\label{lsv-pf}  
\vspace{-10pt}
\end{figure*} 

In Figure~\ref{lsv-pf}(b), we show the variation of PF with the source luminosity, in the energy ranges 1--10\kev, 10--30\kev \& 30--150\kev as obtained respectively from LE, ME and HE pulse profiles. Particularly, in hard X-rays (\ie in 30--150\kev) we see a turnover of PF variation with luminosity twice. The first turnover is at $L_{x}\sim1$\lumcgsts and the second turnover is seen near $L_{x}\sim2.1$\lumcgsts. The variation of PF shows a clear dependence on the X-ray luminosity. Previously, \citep{2023MNRAS.526.3637L} have reported the transitional luminosity of $\sim2$\lumcgsts from their study of PF variations in 30--100\kev pulse profiles. In the present work, we have used the HE pulse profiles in 30--150\kev and found that there exists another turn-over point at $L_{x}\sim1$\lumcgsts at which, the PF variation changes from a negative trend at lower luminosity to a positive variations with $L_{x}$ up to $\sim2.1$\lumcgsts (see Fig~\ref{lsv-profiles}(b)). With further increase in source luminosity (\ie $L_{x}>2.1$\lumcgsts, we see slight decline in the PF variation. Within these two transitional luminosity, we find the PF values in all the energy ranges show a positive dependence on luminosity. Interestingly, in Figure~\ref{lsv-profiles}, we observe that, at the transitional luminosity, $L_{x}\leq1$\lumcgsts, the HE pulse profiles are multi-peaked. The two prominent peaks located near pulse phases 0.1--0.2 and 0.6--0.7. With the increase in luminosity, the first peak becomes less prominent and diminishes in size. However the second prominent peak splits into two at $L_{x}\geq2.4$\lumcgsts. At the outburst peak, with the net decrease in luminosity, we see that the pulse profiles evolved to a single peak.

\subsection{Spectral studies}

\subsubsection{Phase-averaged spectroscopy}\label{avg-spec}
The 2023 Type-II outburst in \rxj has been observed by \hxmt at multiple luminosity epochs of the pulsar covering the progress, decline and peak outburst phases of the pulsar emission. These observations provide the broadband X-ray spectra of the pulsar in 1-120 \kev to study the variable spectrum of the pulsar. \citet{2024A&A...689A.316L} have studied the pulsar spectrum during a high- and low-luminosity state of the pulsar using two \hxmt observations (\ie ObsIDs P0514361040, P0514361005 respectively). In the present work, we have studied the pulsar broadband spectrum as it evolved throughout the giant outburst covering many luminosity epochs. We have performed a combined spectral analysis of the LE, ME and HE data from 2-120 \kev for each of the \hxmt observations. The LE spectra from all exposure-IDs pertaining to a single observation ID were combined together with \emph{addascaspec} {\sc FTOOLs}. This has resulted in a higher effective exposure for each observation used for phase-averaged spectral analysis (Table~\ref{tab:log1}). Similar procedure has been carried out for ME and HE spectra. We have used the LE spectra in 2--10\kev, ME spectra in 10--30\kev and the HE spectra in 30--120\kev for the broadband spectral analysis. At first, we have considered the commonly used phenomenological continuum models to describe the high-energy spectra of X-ray pulsars. These are namely: (a) a cutoff power law (\cpl) model, (b) a power-law model with an exponential cutoff at high energy (\emph{powerlaw*highecut}), (c) a combination of two power laws with a negative and positive spectral index with an exponential cutoff in energy (\emph{NPEX}) \citep{1995AAS...18710403M}. To account for the Galactic interstellar absorption along the line of sight, we apply the \emph{TBabs} absorption model component to these continuum models with the elemental abundances set to \emph{wilm} \citep{2000ApJ...542..914W} and cross sections set to \emph{vern} \citep{1996ApJ...465..487V}. We let the hydrogen column density \nh to vary freely while spectral fitting. However, we note that the Galactic H~I density in the direction of the \rxj is $\sim0.6\times10^{22}$\pcm \citep{2016A&A...594A.116H}. 

\begin{figure}[!t]
 \begin{center}$
 \begin{array}{ll}
 \includegraphics[height=2.5in, width=3.5in, angle=-90]{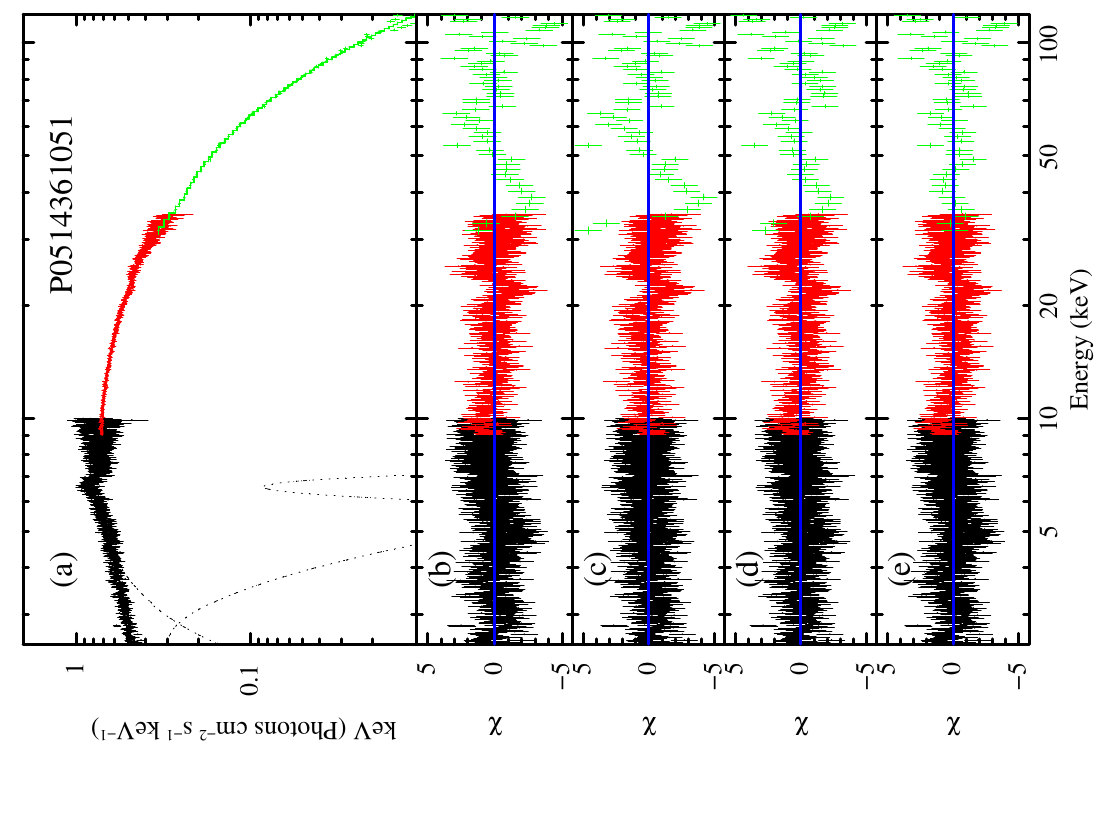} &
 \end{array}$
 \end{center}
 \vspace{-10pt}
\caption{\small The unfolded phase averaged spectra along with best fitted  continuum model shown in the figure in inset (a). The spectral residuals obtained with various phenomenological models such as : (b) an absorbed \cpl\  model, (c) an absorbed power law model with a high-energy exponential cutoff, (d) an absorbed two component power law model with negative and positive exponential cutoff, (e) The best-fitted absorbed cutoff power law model with a cyclotron absorption line component in hard X-rays.   The absorption feature in hard X-rays is fitted with a multiplicative cyclotron line absorption component (\cyclabs\ in \xspec ). The error-bars represent 1$\sigma$ uncertainties. In the inset (a), the observation ID is shown in the right side of the panel. Corresponding epochs of the \hxmt observations can be seen from Table~\ref{tab:log1} in the Appendix.}
\label{lsv-spec0}
\end{figure}

We have performed pulse phase-averaged spectral studies of the pulsar with different spectral continuum models as stated above in combination with a soft blackbody component, such as (1) \cpl + \bbody, (2) \hcut + \bbody, (3) \npex + \bbody. It is found that, a cutoff power law model with a blackbody component  (\cpl+ \bbody) fits the phase-averaged spectra well in all epochs of the source luminosity. In addition to the soft blackbody component at  $kT_{BB1}\sim0.3$\kev to better fit the low-energy spectra in most of the observations, we have also used another additional blackbody component of $kT_{BB2}\sim2.5$\kev during the declining part of the outburst, to explain the broadband spectra. Apart from these two additional blackbody components, we have also included a Gaussian component near 6.4~\kev to account for neutral and fluorescence $Fe~K\alpha$ line emission. The presence of 6.4\kev line in the outburst spectra is also noted from \nicer observations \citep{2023MNRAS.526..771M}. In hard X-rays near 30-40\kev, we find a broad absorption feature in the spectra, regardless of the model combination used as described above. The spectral residuals seen at these energies with the above model combinations is shown in Figure~\ref{lsv-spec0} (\ie see panels b,c,d) considering the example of phase-averaged spectra of \hxmt observation ID P0514361051. This has been explained with an absorption component: \cyclabs\ (in \xspec) thought to be a cyclotron resonance scattering feature \citep{2019A&A...622A..61S} seen sometimes in the spectra of bright X-ray pulsars. 
The expression for \cyclabs\ model in \xspec is:
\begin{equation}
\exp\left(\frac{-\tau_\mathrm{cycl}(E/E_\mathrm{cycl})^2\sigma_\mathrm{cycl}^2}
{(E-E_\mathrm{cycl})^2+\sigma_\mathrm{cycl}^2}\right),
\end{equation}

\noindent where the cyclotron line central energy, width and depth are denoted as $E_\mathrm{cycl}$, $\sigma_\mathrm{cycl}$, and $\tau_\mathrm{cycl}$ respectively (\cite{1990Natur.346..250M}).

\begin{figure}[!t]
 \begin{center}$
 \begin{array}{ll}
 \includegraphics[height=7.0in, width=4.3in, angle=0]{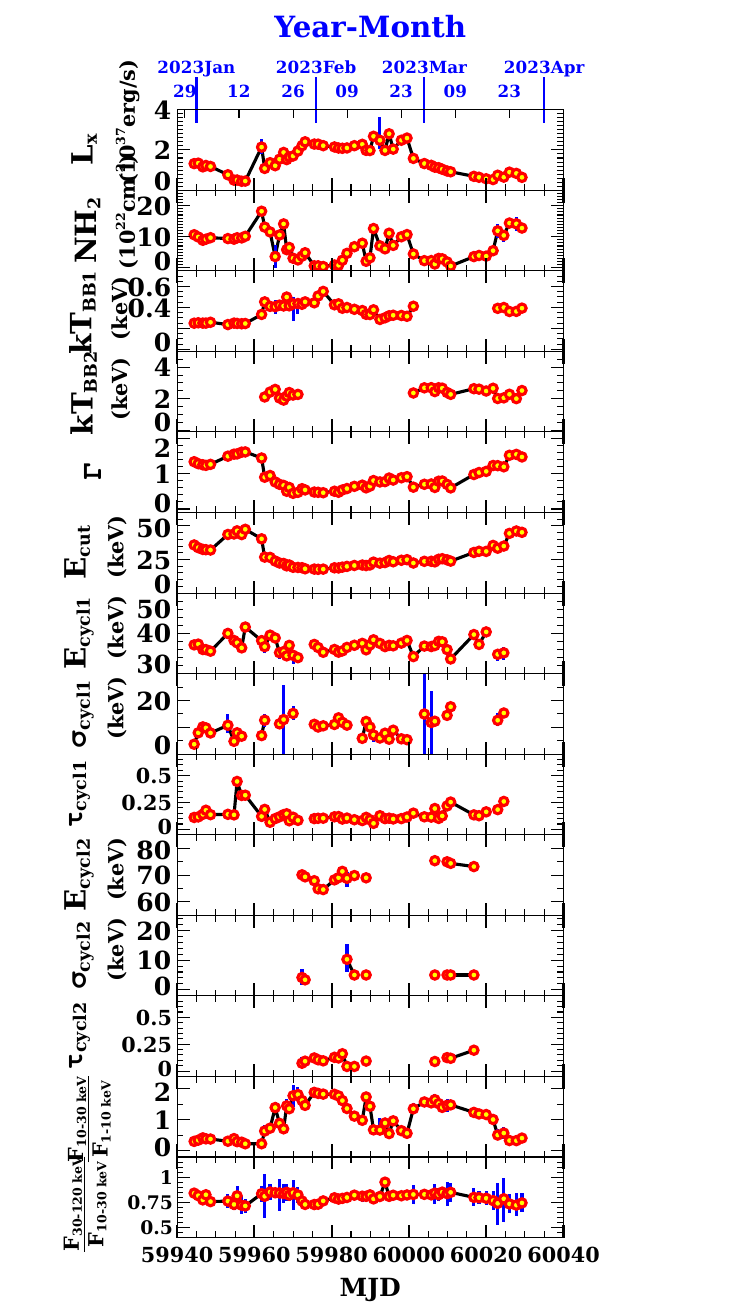} &
 \end{array}$
 \end{center}
\vspace{-15pt} 
\caption{\small Variation of spectral parameters of \rxj at various luminosity epochs of the giant outburst. An absorbed cutoff power-law model (\ie \cpl) with additional components like: 1) a Gaussian component at 6.4\kev, 2) a cyclotron absorption component varying in ~33.6-41.6\kev, 3) a soft black-body component near 0.3\kev, and sometimes a second blackbody component near 2.5\kev has been used to obtain the best-fitted continuum model.}
\vspace{-10pt}
\label{lsv-phase-avg}
\end{figure}

The presence of cyclotron line in the spectra near $\sim30$\kev has previously reported in \citet{2012MNRAS.421.2407T} during a low luminosity state of the pulsar. Recently, however \citet{2023MNRAS.524.5213S} have used two-component models like {\bbodyrad+ \cpl} and {\comptt+ \comptt} to explain the \nustar observed broadband spectra of \rxj in a bright state without the need of including CRSF near 30\kev. Apart from the detection of fundamental cyclotron line near $\sim30$\kev, we have also detected a first harmonic of the cyclotron line varying in energy $\sim64.6-75.3$\kev of the broad-band spectra near the peak and declining phases of the outburst. However, it can be seen from the Figure~\ref{lsv-phase-avg} that although the line width of these cyclotron lines are quite wide (i.e $\sigma_{cycl1}> 5$\kev,  $\sigma_{cycl2}> 5$\kev in most of the observed epochs during the giant outburst), the line depth value, $\tau_{cycl1}$ and $\tau_{cycl2}$  are shallow (\ie < 0.5) signifying the weak detection of these lines.

For all the \hxmt observations of \rxj covering the entire giant outburst (Figure~\ref{lsv-swift-bat}), we have performed the phase-averaged spectral analysis with the best fitted \cpl+\bbody\ model along with the additional model components. The variation of spectral parameters with source X-ray luminosity is shown in Figure~\ref{lsv-phase-avg}. The unfolded phase-averaged spectra obtained with best fitted continuum models and respective residuals are shown in Figure~\ref{lsv-spec0}. It can be seen from Figure~\ref{lsv-phase-avg} that the low temperature blackbody component, \kt varies in the range 0.24 -- 0.55 \kev, and the high-temperature blackbody component, \ktt varies in the range of 1.9 -- 2.7 \kev. The second blackbody component is seen at source luminosity $\leq2$\lumcgsts. During the declining phase of the giant outburst, we find the presence of the high-temperature blackbody component varying in 2--2.7 \kev, while the soft blackbody component is absent in the spectral continuum. The variation in power-law photon index ($\Gamma$) shows a negative trend with increase in source luminosity up to $\leq2$\lumcgsts, whereas with further increase in source luminosity, $\Gamma$ shows a positive trend up to the maximum source luminosity as seen during the giant outburst. From the variation of cutoff energy ($E_{cut}$) with source luminosity we see a similar variation as seen in case of $\Gamma$.  For luminosity $\leq2$\lumcgsts, the variation in $E_{cut}$ shows a decreasing trend from a maximum value of 47.5\kev at low source luminosity of $(0.46\pm0.10)$\lumcgsts to a minimum cutoff energy, $E_{cut}$ value of 17.6 \kev near the outburst peak luminosity of $(2.29\pm0.02)$\lumcgsts. However with further increase in luminosity as shown in Figure~\ref{lsv-phase-avg-2}, we see a positive trend of $E_{cut}$ with $L_{x}>2$\lumcgsts.  

In Figure~\ref{lsv-phase-avg}, we also show the variation of two flux ratios with the progress of the giant outburst. We define the first flux ratio in soft X-rays (\ie HR1) as the ratio of X-ray flux in 10--30\kev to that of X-ray flux in 1--10 \kev, the second flux ratio in hard X-rays (HR2) as ratio of flux in 30--120\kev to that of flux in 10--30 \kev. Interestingly, we find that, HR1 increases with source luminosity during the progress of the outburst. However at the outburst peak (\ie between MJDs 59976.5 -- 60001.1) as can be seen from Figure~\ref{lsv-phase-avg}, we see a decline of HR1 with increase in source luminosity. However, beyond MJD 60001.1 as the outburst was declining, we find that, HR1 shows a decreasing trend with decrease in source luminosity. We note that, the onset of decline in HR1 values occurred when the luminosity was ($2.29\pm0.02$)\lumcgsts. The source outburst luminosity during the dip of HR1 variation at the outburst peak is $\geq2$\lumcgsts. In the variation of HR2 with outburst luminosity and throughout the outburst duration, we do not observe any such abrupt changes. Rather, the HR2 values vary within 0.72-0.95, mostly showing no significant variation either with the outburst luminosity or with the progress of the giant outburst. 

\subsubsection{Phase-resolved spectroscopy}\label{sec:empirical-phase-res}

The interesting variation of phase averaged spectral parameters with source outburst luminosity, prompted us to explore further the spectral parameter variations with pulse-phases of the pulsar. For this, we have carried out pulse-phase resolved spectral studies at three different luminosity epochs of the pulsar. Two of the phase-resolved studies are performed at similar luminosity of $(1.22\pm0.09)$\lumcgsts and $(1.24\pm0.05)$\lumcgsts in 1--120\kev respectively during the progress (\ie at MJD 59965.37 with ObsID: P0514361025) and declining phase (\ie at MJD 60005.77 with ObsID: P0514361057). The pulsar spectra at these epochs has been divided into 10 pulse-phase bins. Spectra from LE, ME and HE for each pulse-phase bin are fitted simultaneously with the best-fitted spectral model for these observations as obtained during the phase-averaged spectral studies. The third phase-resolved spectral study is performed for the pulsar observation taken at the peak of the giant outburst (\ie near MJD 59977.8 with ObsID: P0514361038) where the net luminosity of the pulsar was noted to be ($2.21\pm0.02$)\lumcgsts. The pulsar spectra at this epoch obtained from HE, ME and LE detectors have been divided into 20 pulse-phase bins. 

\begin{figure*}
 \begin{center}$
 \begin{array}{cc}
 \includegraphics[height=4in, width=5.5in, angle=0]{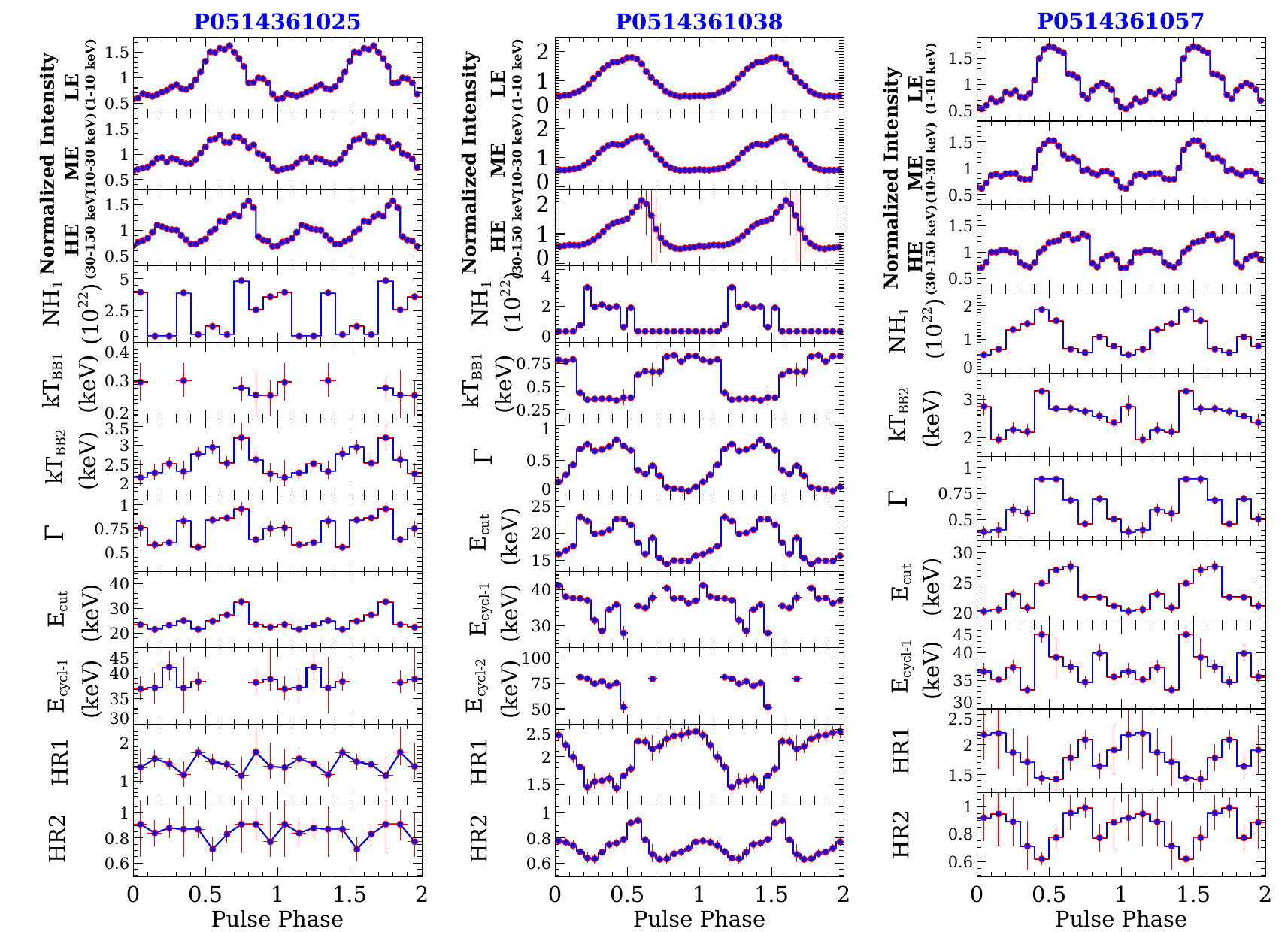} & 
 \end{array}$
 \end{center}
 \vspace{-15pt}
\caption{\small Variation of spectral parameters with pulse phases at three luminosity epochs of the pulsar using an absorbed cutoff powerlaw (\ie \cpl) model along with blackbody components. The absorption feature in hard X-rays is fitted with a cyclotron line absorption component (\emph{cyclabs}). The error-bars represent 1$\sigma$ uncertainties. The observation IDs are shown on the top of each panel. Corresponding epochs of the \hxmt observations can be seen from Table~\ref{tab:log1} in the Appendix.}
\label{lsv-phase-res-hcut}
\end{figure*} 

For the phase-resolved spectral fit of the ObsID: P0514361025, we have used the best fit absorbed \cpl\ model with two blackbody components (with \kt$\sim0.4$\kev and \ktt$\sim2.6$\kev), a Gaussian component at $\sim6.4$\kev, and a cyclotron absorption line component ($E_{cycl}\sim38.85$\kev). For the phase-resolved spectral studies at the peak of the giant outburst, we have considered spectra from the ObsID: P0514361038. The best-fit phase averaged spectral continuum for this epoch is an absorbed \cpl\ continuum model with a soft blackbody component at $\sim0.55$\kev and two cyclotron line components with line energies at $\sim35.3$\kev and $64.6$\kev respectively. In the declining phase of the giant outburst, we have carried out pulse phase-resolved spectral studies from the \hxmt observation having ObsID: P0514361057, where the best-fit phase averaged spectral model consists of an absorbed \cpl\ model with a single blackbody component at 2.7\kev and a cyclotron absorption line component at energy $\sim36.7$\kev. In Figure~\ref{lsv-phase-res-hcut} we show the spectral parameter variations obtained with pulse phases during these three luminosity epochs along with the pulse profiles in energy bands 1--10\kev, 10--30\kev, 30--120\kev. 

In ObsID: P0514361025, the soft blackbody emission component (\ie \kt) varies within (0.26--0.30) \kev over pulse phases (Figure~\ref{lsv-phase-res-hcut}). And it is not seen at the peak phases of pulse profile (\ie in 0.4 -- 0.7 pulse-phase bin). However, the second blackbody component (\ie \ktt) is seen across all pulse-phase bins and varies within (2.16 -- 3.2) \kev . And the variation in \ktt is found to be in phase with the LE and ME pulse profiles. The CRSF at $38.85^{+0.62}_{-0.58}$\kev in the phase-averaged spectra is found to vary in (36.9 -- 41.2)\kev over pulse phases of the pulsar except the peak pulse phases, where it is absent in the broadband pulsar spectra. In case of the phase-resolved spectra of ObsID: P0514361038 (at outburst peak), we found the pulse profiles to be single peaked and evolve with energy. The column density \nh was marginally high at pulse phases 0.2--0.6 compared to the Galactic column density along the source direction. During these pulse phases of high intensity, \kt varied at lower values of $\sim0.35$\kev, while in the low intensity phases, \kt increased to higher values of $\sim0.8$\kev. In other words, the variation in \kt remained out-of-phase with the pulse profile. Similar out-of-phase variation with pulse profiles is seen in case of HR1 variation with pulse phases. However the variation of $E_{cut}$ and $\Gamma$ with pulse phases was found to be similar and in phase with the pulse-profile variations. The maximum value of $E_{cut}$ was 22.97\kev and $\Gamma$ is 1.63 at maximum intensity of pulse profiles. Whereas the minimum  values of $E_{cut}\sim14.34$ and $\Gamma\sim0.11$ were seen at $0.775\pm0.025$ pulse phase where the HE, ME and LE pulse profiles are also seen to be at low intensity phases. Interestingly, we found the presence of double cyclotron lines  in the pulsar phase-resolved spectra that vary with pulse phases. The first CRSF feature was found at $E_{cycl-1}=35.3\pm0.4$\kev in the phase-averaged spectra with $\sigma_{cycl-1}=10.6\pm0.7$\kev and a narrow line width of $\delta_{cycl-1}=0.10\pm0.01$, which varied with pulse phases, with $E_{cycl-1}\sim$ (28.05--41.21)\kev. The second cyclotron line component is found to vary from ($51.77 - 80.96$)\kev. Moreover, it is observed in 0.2-0.7 pulse phases while undetected during the minimum intensities phases of the pulse profiles. During the giant outburst peak of the pulsar, the variations in cyclotron line energies show a strong pulse-phase dependence.  

In order to check the pulse-phase dependence of spectral parameters during the decline of the giant outburst and any similarities in the spectral properties as observed during the progress of the giant outburst, we have performed phase-resolved spectral studies of the pulsar from the broadband spectra obtained from the \hxmt ObsID: P0514361057. We note that, the 1-120\kev unabsorbed flux (and hence luminosity) of the pulsar obtained during the progress of the outburst (ObsID: P0514361025) and during the decline of the outburst (ObsID: P0514361057) are similar. Therefore it is interesting to check the spectral parameter variations during the decline phase of the outburst from the pulse-phase resolved spectral studies of \hxmt observation with ObsID: P0514361057. During this observation, the soft blackbody component \kt which varied with pulse phases during the progress and at the peak of the giant outburst, was not needed in the broadband spectral fitting with the empirical \cpl\ model. However, a high temperature blackbody component \ktt was required in the phase-resolved spectral fit, and it is found to vary with pulse phases of the pulsar in the energies (1.95--3.23)\kev. The power-law spectral index $\Gamma$ also varies with pulse phases in the range (0.39--0.89) which is higher at high intensity phases of the pulse profiles and vice versa. The cutoff energy ($E_{cut}$) was also found to vary with pulse phases similarly in the range (20.24--27.72)\kev, which is low at low-intensity pulse phases and reaching higher values at pulse phases where pulse profiles show maximum intensity. The cyclotron line energy ($E_{cycl-1}$) was seen to vary with pulse phases in the range (33.34--43.24) that is marginally higher at the primary and secondary sub-peaks (\ie in 0.4--0.5  \&  0.8--0.9 pulse phases) of the pulse profiles. The variations in HR1 and HR2 values across pulse phases show similar patterns of variability, that is lower at high intensity phases of pulse profiles and vice versa. 

\subsection{Results from Comptonization model:}
\begin{figure*}
    \centering
    \includegraphics[height=5.6 cm,angle=-90]{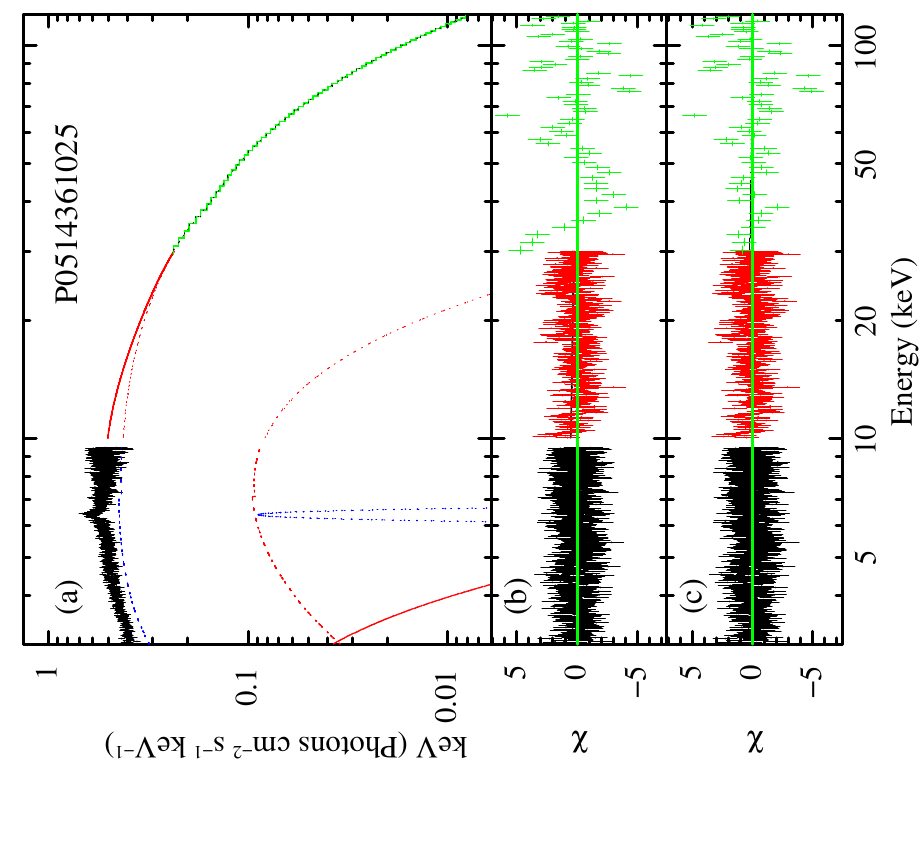}
    \includegraphics[height=5.6 cm,angle=-90]{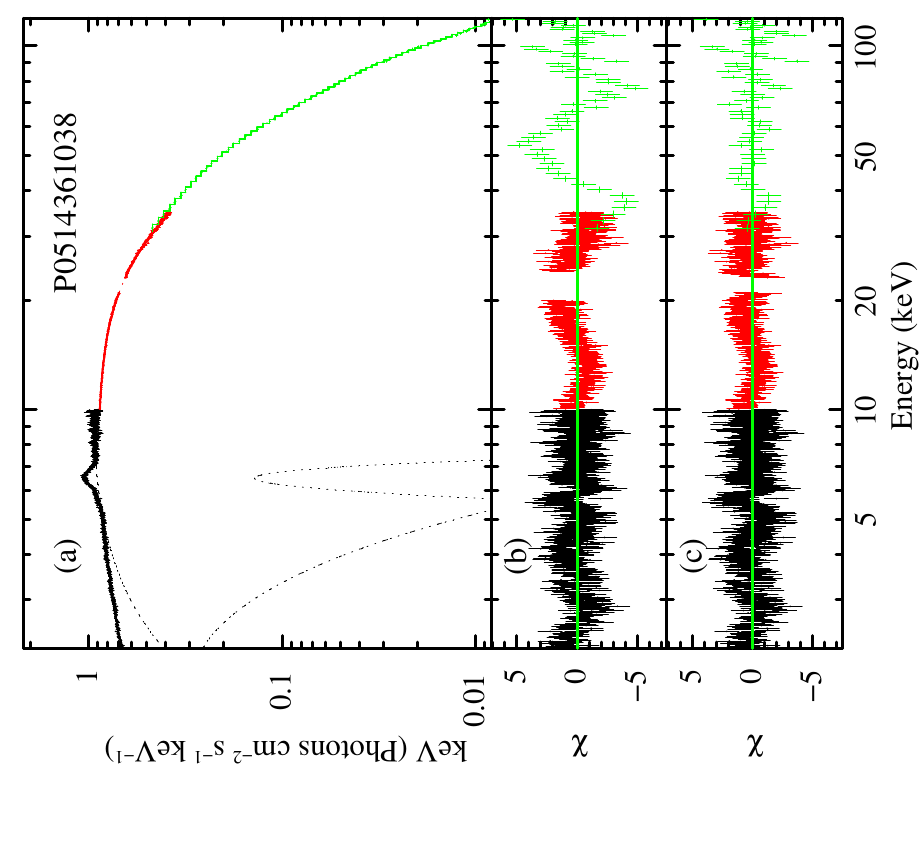} 
    \includegraphics[height=5.6 cm,angle=-90]{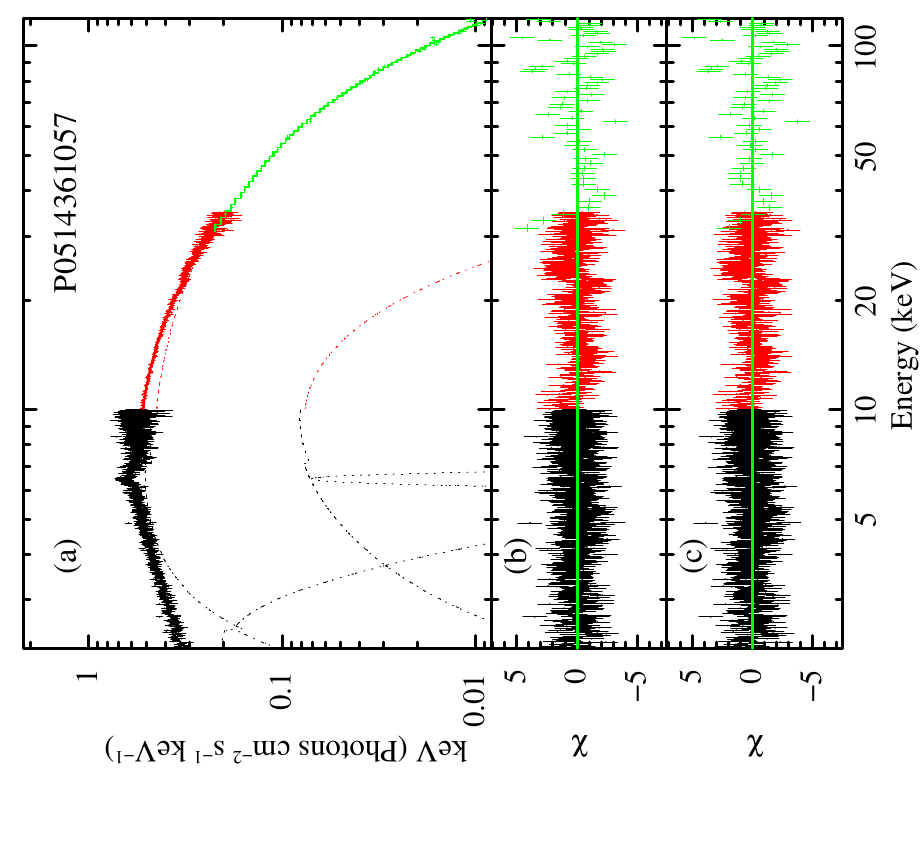} \\
    \caption{\small (a): The unfolded spectra of \rxj in 2-120\kev obtained from three example observations during the late 2022-2023 giant outburst. The spectrum in the left is obtained from ObsID:~P0514361025 (near MJD 59965.37) before the outburst peak. The spectrum shown in the middle is obtained from the ObsID:~P0514361038 taken at the peak of the outburst (MJD: 59977.804). The spectrum shown in the right side is obtained from the declining phase of the giant outburst at MJD 600005.769 (ObsID:~P0514361057). The spectral residuals shown in the panels (b) in each figure are obtained from the BW model without any cyclotron line component. Whereas the residuals shown in panels (c) are for best-fitted BW model requiring a cyclotron line component (in all of the three mentioned observations) and a second cyclotron line component (for the ObsID: P0514361038) in hard X-rays . A Gaussian component near 6.4\kev is added to the spectra to account for the emission from neutron Fe~$K\alpha$ line near to the pulsar in \rxj.}
    \label{lsv-bw-spec}
    \vspace{-10pt}
\end{figure*}

In the previous sections~\ref{avg-spec} \& \ref{sec:empirical-phase-res}, the strong variability of the pulsar spectrum has been explained with a phenomenological model. However, it lacks a complete physical description of the pulsar accretion column emission during the bright outburst. Given the pulsar observed with \hxmt at large luminosity ranges, it is imperative to seek a physical description of the accretion column emission of the pulsar using these observations. Therefore we have used a physics-based Comptonization model \citep{2007ApJ...654..435B} to the explain the pulsar spectra at its broad luminosity epochs. The Comptonization model (\ie hereafter BW07 model), could explain the pulsar accretion column emission spectra considering energization of the thermal and bulk Comptonization of seed photons. Wherein it is assumed that the seed photons produced in the cylindrical accretion column are primarily due to the bremsstrahlung, cyclotron and blackbody emission processes in the accretion plasma \citep{2007ApJ...654..435B, 2022ApJ...939...67B}.  The BW07 model has been useful in explaining the spectra of bright and highly magnetized X-ray pulsars, like \fuo \citep{2009A&A...498..825F}, \exo \citep{Epili2017,2024ApJ...969..107Y}, \herx \citep{2016ApJ...831..194W,2007ApJ...654..435B}, \cenx and \lmcx \citep{2007ApJ...654..435B}. The \xspec incorporation of this model (as \bwcycl) and its usage has been illustrated in \citet{2009A&A...498..825F}. Among the BW model parameters, we fix the mass and radius of the neutron star (\ie $M_{NS}=1.4$\msol and $R_{NS}=10$\rsol), and the distance is fixed at values of 2.44\kpc \citep{2021AJ....161..147B}. The remaining free parameters of the model are: (1) the mass accretion rate (\mdot), (2) the electron temperature $T_{e}$, (3) the column radius $r_{0}$, (4) strength of the NS magnetic field, $B_{12}$ in units of $10^{12}$~G, (5) the dimensionless photon diffusion parameter $\xi$, and (6) the Comptonization parameter $\delta$. Among these, the parameters $\xi$ signifies the relative importance of the timescale for the radiation to diffuse through accretion column walls to the dynamical time scale for the gas to accrete onto the NS. it is defined in equation (26) of \cite{2007ApJ...654..435B} as :   
\begin{equation}
\xi=\frac{\pi r_0 m_p c}{\dot m \left(\sigpar
\sigperp\right)^{1/2}},
\label{eq:eq1}
\end{equation}
where $m_p$ denotes the proton mass, $c$ is the speed of light, and $\sigpar$ and $\sigperp$ respectively represent the mean scattering cross section for photons propagating parallel and perpendicular to the magnetic field. The ratio of corresponding Compton parameters $y$ for the bulk and thermal Comptonization $\delta$ is expressed as (see equation (98) of \citealt{2007ApJ...654..435B}):
\begin{equation} 
\frac{\delta}{4}=\frac{y_\mathrm{bulk}}{y_\mathrm{thermal}}.
\label{eq:eq2}
\end{equation}
%
\begin{figure}
 \begin{center}
   \includegraphics[height=4.5in, width=3.3in, angle=0]{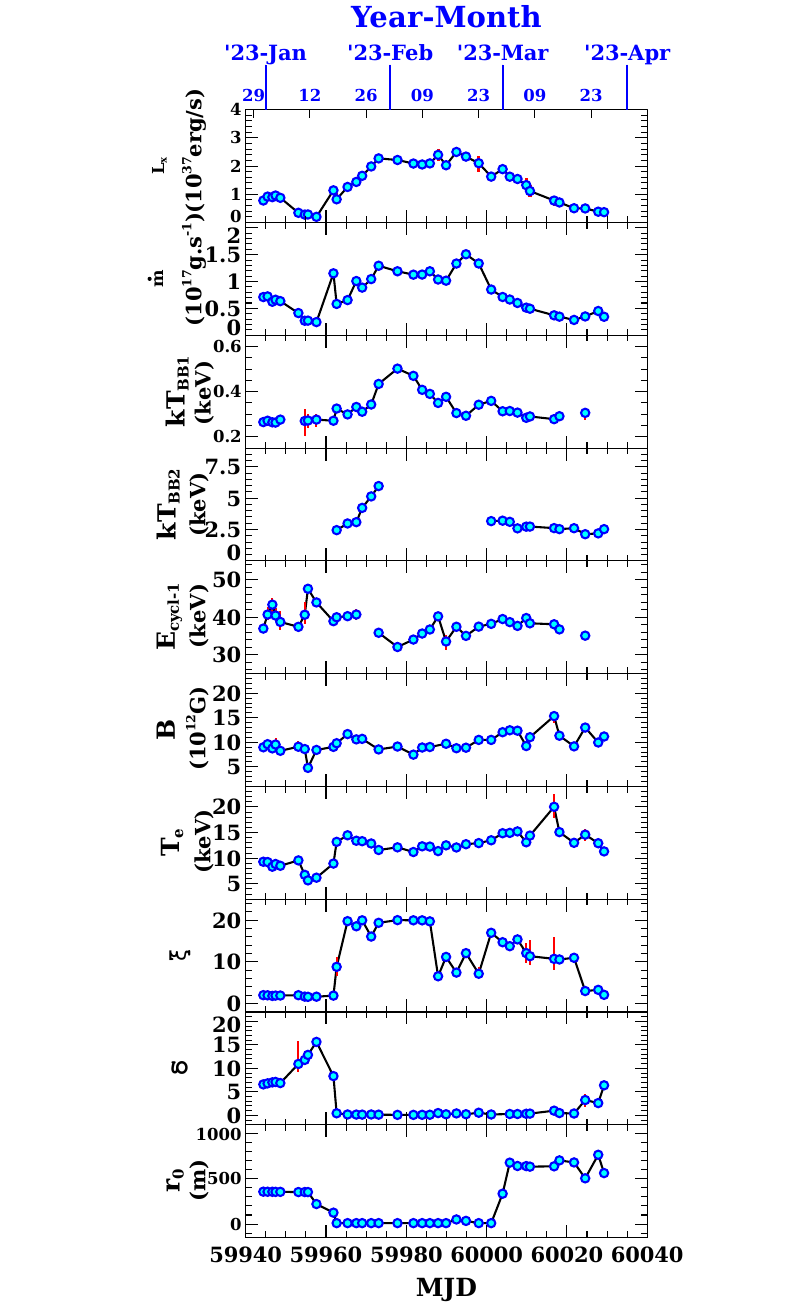} 
 \end{center}
 \vspace{-15pt}
 \caption{\small The best-fitted phase averaged spectral parameters using \emph{bwcycl} model obtained at different epochs of the the giant outburst in \rxj as observed with \hxmt. The error bars represent 1$\sigma$ uncertainties.}
 \label{lsv-avg-bw}
\end{figure}  
 
Among the other free parameters, during the spectral fitting procedure, we constrain the mass accretion rate $\dot m$ by assuming that, the total energy integrated luminosity, $L_{x}$ of the emergent spectrum of the pulsar is equal to the net accretion luminosity, $L_{x}=L_{acc}=GM_{NS}\mdot/R_{NS}$ of the pulsar observed in 1-120 \kev.  We have used the phase-averaged best-fit empirical \cpl\  model along with the additional components, as described in section\ref{avg-spec} to estimate the flux in 1-120 \kev (see Appendix~\ref{tab:log1}). The mass accretion rate (\mdot) is strongly correlated with $r_{0}$, then after obtaining a best fit with the BW07 model, we freeze the parameter values of $r_0$ and \mdot as suggested in \citep{2009A&A...498..825F}. This ensures a stable fit for other model parameters such as $B_{12}$, $T_e$, $\xi$ and $\delta$. 
The absorption model component, \tbabs\  has been used with the scattering cross-sections set to \texttt{vern} and abundances set to \texttt{wilm} \citep{2000ApJ...542..914W}, to account for the Galactic absorption in the direction of \rxj. To better fit the spectra at low energy X-rays, we require a soft blackbody component \bbody\  of temperature \kt, a Gaussian component at $\sim6.4$\kev. A high-temperature second blackbody component of temperature \ktt, was also required while fitting the pulsar spectra observed during the progress and declining phases of the giant outburst (similar to \cpl\ model in Section~\ref{avg-spec}). For the other additional model components of the BW model, in hard X-rays we have used an absorption component (namely, \cyclabs) to include cyclotron resonance scattering features (CRSFs) seen in the spectra. The line energy of the CRSF component is found to be varying in ($\sim32.1-47.6$)\kev during the outburst. The line width of this feature has been fixed to 5\kev as it could not be constrained during the spectral fit. Apart from this absorption feature, during the peak and declining phases of the giant outburst (\ie MJD: 59973.11 -- 59994.87), we found a second absorption feature in hard X-rays. Therefore we have used another \cyclabs\ component in the broadband spectral fitting for these observations. The energy of the second CRSF component is found to be (63.8 -- 77.7)\kev. In Figure~\ref{lsv-bw-spec}, we have shown the unfolded phase-averaged spectra of three \hxmt observations (\ie from ObsdID(s): P0514361025, P0514361038 and P0514361057) with the best fitted model and residuals. The physical parameter variation of the best-fit BW07 model obtained across the giant outburst at 37 different luminosity epochs of the pulsar \rxj is shown in Figure~\ref{lsv-avg-bw}.

Using the BW model we could estimate the pulsar magnetic field (\ie $B_{12}$ in units of $10^{12}G$) of the pulsar as an independent parameter. Throughout the outburst, we found that, the estimated pulsar magnetic field is very high and varies in the range $(\sim4.8-15.4)\times10^{12}G$. During the bright phases of the outburst (\ie at $L_{x}\geq2$\lumcgsts), $B_{12}<10$, whereas during the declining phases as the net X-ray luminosity of the pulsar decreases to lesser values than $2$\lumcgsts down to $\sim0.37$\lumcgsts, we find a higher estimate of the NS magnetic field reaching $(15.3\pm1.4)\times10^{12}$G  Figure~\ref{lsv-avg-bw}. The mass accretion rate,\ \mdot$_{17}$ (\ie \mdot in units of $10^{17}gs^{-1}$) during the outburst varied in the range $(\sim0.25-1.51)$ reaching higher values with increase of source luminosity. At the outburst epoch MJD:59994.86, the pulsar in \rxj accreted at a maximum rate of \mdot$_{17,max}\sim1.51$. However, it reduced to lower values of $\sim0.29$, during the declining phase of the outburst at MJD: 60021.773. The temperature of additional soft blackbody component, \kt in the range of $(\sim0.26-0.50)$\kev increases with the source luminosity. Noticeably, near the peak    episodes, we see an abrupt decline of the \kt values to $0.29\pm0.01$\kev which later increases to ~0.36\kev before further decreasing to lower values of $0.28\pm0.02$ as the outburst declines beyond MJD 60017. Similar to the variations seen with the empirical best-fit \cpl\ model, in explaining the phase-averaged spectra (\ie as in section\ref{avg-spec}), we also see an evolution of additional high-temperature blackbody component (of temperature \ktt) along with BW model parameters. The second \bbody\ component was required in the spectral fitting during the progress of outburst (\ie during MJD: 59962.65 -- 59973.11) increasing in the energy, \ktt (2.45--5.97)\kev and during the declining phase of the outburst (\ie during MJD: 60001.14 -- 60029.25) with a decreasing trend in \ktt from (3.17--2.53)\kev. We find that, the second \bbody\ component was not needed during the outburst peak episodes, similar to the results seen from empirical best-fit model to the phase averaged outburst spectra of \rxj. 

In hard X-rays above 30\kev, we have used an absorption component, namely \cyclabs\ to account for the CRSF seen in the spectra varying in energy, $E_{cycl1}$:(32.08 -- 47.56)\kev with the outburst epochs of the pulsar and showing no particular dependence on outburst source luminosity. However, as can be seen from the Figure~\ref{lsv-avg-bw-2}, between the source outburst luminosity, $(1-2)$\lumcgsts, we see that,  $E_{cycl1}$ is negatively correlated with the net source X-ray flux (or luminosity). Apart from the first cyclotron line component, $E_{cycl1}$, we have also detected a second cyclotron line component in hard X-rays during the spectral fit with BW model to the broadband spectra obtained from declining phases of the giant outburst (\ie during MJD 59973.11 - 60009.87) . The line energy of this second CRSF component varied in ($\sim63.8-77.7$)\kev. The line depth of the first CRSF was found varying in (0.05 -- 0.29) during the entire outburst signifying the weak detection of this absorption feature. Whereas the line width during most of the epochs is fixed at 5\kev to obtain better constraints on the error estimates of other spectral parameters.

Among the other BW07 model parameters, the plasma temperature ($kT_e$), in other words the temperature of the electrons in the optically thin regions above the thermal mound of NS, varies highly, reaching higher values of 20 \kev during the declining phases, whereas during the initial phases, $T_e\sim5.7$\kev. The average plasma temperature during the outburst peak episodes (between MJD:59961--60010) remained at values $<13$\kev. It shows there is a substantial cooling of plasma seen  at the high source luminosity phases of the outburst. This is also evident from the variations of $T_e$ vs $L_x$ as shown in Figure~\ref{lsv-avg-bw-2}. Such a scenario may occur near pulsar accretion column, when pulsar emission is close to its \emph{critical luminosity} regime. During which, there can be a change of emission beam pattern of the pulsar from a pure pencil beam pattern at low X-ray luminosity to a mixture of pencil beam and fan-beam emission pattern. Therefore the hard X-ray photons escape the accretion column through side walls, leading to a decrease in the accretion plasma temperature near the emission regime of critical luminosity.  This has been further strengthened from the variation in the dimensionless parameter, $\xi$ with $L_x$ during the peak of the giant outburst. As defined in equation~\ref{eq:eq1}, the parameter $\xi$ shows the relative importance of the timescale for the photons to escape through the accretion column walls to the dynamical scale for the accretion of the gas onto the NS. 

In terms of accretion column height, it is expressed in equation 109 of \citep{2007ApJ...654..435B}, as ratio of "trapping altitude" ($z_{trap}$) for the photons to diffuse vertically in the accretion column to the distance from the NS surface to the sonic point, $z_{sp}$: \(\frac{z_{trap}}{z_{sp}}\sim2\xi\). From the variation of $\xi$ with the luminosity, we find that, higher values of $\xi>10$, indicating higher values of trapping altitudes compared to $z_{sp}$, leading to photon penetration into the higher regions of above the sonic point in the accretion column. From the variation of $\delta$ parameter with the progress of the outburst, we find that, it is varying at its lower values within $\leq1$ throughout the giant outburst. However, during the onset and declining phases of outburst when the accretion luminosity $L_x<1$, the $\delta$ was seen at values $>>1$. As per equation~\ref{eq:eq2}, this implies that, it is the thermal Comptonization as the dominant mechanism in shaping the pulsar accretion column emission, where the photon energization proceeds through thermal scattering of plasma electrons. The size of column emission region ($r_0$) during the outburst peak episodes (\ie between MJD: 59962--60001) is found to be $<50m$, indicating most of the Comptonized column emission is seen from the base of accretion column. However, with the decline of the outburst, column emission size rises to higher values (\ie $r_0>500m$), implying that the net Comptonized emission is seen at higher heights of the accretion column, with the bulk Comptonization becoming the leading mechanism (\ie due to increase in $\delta$) in energizing the X-ray photons. We have also studied the phase-resolved spectra of the pulsar using the BW model at three luminosity epochs which is elaborated in Appendix~\ref{phase-res-BW}.

\section{Discussions}

\subsection{Spin period variations}
The 2022-2023 giant X-ray outburst of Be/X-ray binary (BeXB) pulsar in \rxj has been extensively observed by \hxmt\ spanning several observation epochs during MJDs: 59944 -- 60029. The net X-ray luminosity of the pulsar in 1--120\kev during the outburst spanned two orders of magnitude.  It is found to be in the range of $4.5\times10^{36}$\lumcgs -- 2.8\lumcgsts, where a distance to the source is considered to be 2.44\kpc \citep{2021AJ....161..147B}. However, from earlier observations during the periods of low X-ray activity, it is reported that the net luminosity of the pulsar could reach lower levels of $10^{34}-10^{35}$\lumcgs \citep{2012A&A...539A..82L, 1999MNRAS.306..100R}. A significant spin-up of the pulsar was observed during the outburst in 2022-23 as noted by \citet{2023MNRAS.526.3637L} from the measured spin-period of the pulsar in (30--100)\kev with \hxmt  and in (2--10)\kev with \nicer. From the measured, spin period of the pulsar in (10--30)\kev during the outburst using ME light curves, we find it to be spinning up from ($208.03\pm1.48$)s at the start of the outburst (near MJD 59944.40) to ($204.80\pm0.19$)s near MJD 60029.28 as the outburst declines. The average spin-up rate of the pulsar during its 2022-2023 Giant outburst is found to be $\dot{P}=-(4.4\pm1.7)\times10^{-7} s~s^{-1}$. The present spin-up of the pulsar is comparatively higher by a two orders in magnitude to the average spin-down rate of $\dot{P}=(6.4\pm1.3)\times10^{-9} s~s^{-1}$ seen during the last decade \citep{2012A&A...539A..82L}. 

\subsection{Pulse profiles at similar luminosity and absorption dips}

In this paper, we have studied the variation in pulse-profiles and pulse-fractions across many epochs of the giant outburst. The pulse-profiles of the pulsar found to be highly variable with energy and source luminosity. The energy dependence of the pulse-profiles in \rxj during the  outburst reported earlier \citep{2024MNRAS.534.1028S, 2023MNRAS.526.3637L, 2023MNRAS.526..771M, 2023MNRAS.524.5213S} from \astrosat, \hxmt, \nustar and \nicer observations at different luminosity epochs of the pulsar. In this work, we have compared the pulse profiles in three energy ranges obtained at similar source luminosity with \hxmt during the rising and declining phases of the outburst (see Figure~\ref{lsv-profiles}). We find that the pulse profiles in hard X-rays (30--150\kev) show similar structure and shape at similar values of source luminosity. 

During the declining phases of the outburst, the pulse-profiles at low (1--10\kev) and medium (10--30\kev) energies evolve to be multi peaked, with the appearance of sub-pulses before  and  after the main peak. This is seen in the LE and ME pulse profiles at luminosity $\geq2$\lumcgsts. Appearance of such sub-pulses are also seen reported earlier at source luminosities $\sim6\times10^{36}$\lumcgs (\ie scaled to the recent distance estimate of 2.44\kpc) \citep{2012MNRAS.421.2407T}. At the outburst peak luminosity, we find that the LE and HE pulse profiles become sinusoidal and single peaked. The other prominent feature in the pulse profiles of \rxj is the appearance of an sharp absorption dip like structure at certain pulse phases of the pulsar. It is first reported in \citep{2012PASJ...64...79U} at low energy pulse profiles obtained from RXTE/PCA observations during a outburst of the source in 2010. However during a low luminosity state of the pulsar, such dip features are not seen in the pulse profiles \citep{2012MNRAS.421.2407T}. This speculates some luminosity dependence of the dip structure seen in the profiles. In our present study, we find the dip structure in the low energy, LE(1--10 \kev) pulse-profiles at 0.7 pulse phases during the declining phases of the outburst. The dip structure is found to be absent during the rising phases of the outburst at similar luminosity as shown in Figure~\ref{lsv-profiles}. It is seen up to source luminosity 2.48\lumcgsts. However, during the outburst peak we find the LE pulse profiles to be single peaked with the absence of any dip structures. Instead two sub-pulses before and after the main peak become apparent. Earlier, \citet{2012MNRAS.421.2407T} have reported the appearance of sub-pulses around the main peak in 3--8 \kev, 8--14 \kev pulse profiles in a low luminosity state ($L_{x}\sim1.5\times10^{36}$\lumcgs) and the presence of absorption dip structure after 0.25 phase of the main peak in pulse profiles below 8\kev at high luminosity state (\ie $L_{x}\sim8\times10^{36}$\lumcgs). 
Our timing studies of the pulsar in \rxj shows that the shape of the pulse-profiles remains similar at similar values of source luminosity. This further strengthens fact that the pulse-profiles are strongly luminosity dependent. The structure of the pulsar emission column remains robust during the giant outburst. Although the giant outbursts in X-ray pulsars remains aperiodic and unpredictable, but the strong luminosity dependence of the pulsar clearly shows that, the structure of the accretion column remains unaffected with respect to different outbursts. Another such pulsar which has shown in the past such strong luminosity dependence of pulse-profiles is \exo \citep{Epili2017}.

\subsection{Spectral variability}

To understand the spectral variability during the bright phases of the pulsar emission we have used an empirical continuum model such as an absorbed \cpl\ model along with the additional model components. To have a physical description of accretion column emission in \rxj, we have applied a self consistent bulk and thermal Comptonization model \citep{2007ApJ...654..435B} to all the \hxmt observed epochs of the outburst spectra. This uniquely addresses the changes in the accretion plasma temperature and the scattering cross-sections for the photons propagating parallel and perpendicular to the NS magnetic field. 
BeXBs are generally thought to be transient X-ray binary systems characterized by two types of outburst activity \citep{2011Ap&SS.332....1R}. The peak luminosity during the periodic Type-I outburst reaches $< 10^{37}$\lumcgs, whereas during the aperiodic major Type-II outburst, the source X-ray flux could reach $10^{3}-10^{4}$ times the flux levels seen during quiescence (\ie it could reach luminosity in the range of $10^{37}-10^{38}$\lumcgs). The peak luminosity during the Type-II outbursts in BeXBs sometimes exceed the {\it critical} luminosity emission of the pulsar making them super-critical. Some of the X-ray pulsars reaching the super-critical state of accretion are : \exo\citep{Epili2017, 2024ApJ...969..107Y}, \grojs\citep{2023A&A...669A..38M}, \rxj\citep{2023MNRAS.524.5213S}.  From the \nicer observation of 2022-2023 giant outburst in \rxj, \citealt{2023MNRAS.524.5213S} estimate the \emph{critical} luminosity of $L_{crit}\approx2.8$\lumcgsts from the turn over point noticed in the hardness ratio (7--10 \kev/4--7\kev) versus 1-10 \kev luminosity  variations.

\begin{table*}[!ht]
\tiny
\caption{\label{t7}Computed BW Model Parameters.}
\vspace{-5pt}
\centering
\begin{tabular}{cccccccccccc}
\hline\hline\\
 OBSID      &MJD       &$L_{x}$(\lumcgsts) &$\alpha$ &$\sigpar/\sig$  &$\sigbar/\sig$($\times10^{-4}$)  &$J$($\times10^{7}$) &$\Tmound$($\times10^{7}$K)      &$\vmound/c$   &$\taumound$ &$\taumax$ &$\tautrap$ \\\\
\hline
P0514361017 &59955.509 &$0.28\pm0.04$ &$0.457$   &$3.572 \times 10^{-2}$     &$381.22$                               &$0.69$            &$0.81$    &$0.009$      &$0.019$     &$0.265$             &$1.48$  \\
P0514361022 &59961.859 &$1.14\pm0.08$ &$0.536$   &$0.018 \times 10^{-2}$     &$2.23$                                 &$23.14$           &$2.86$    &$0.019$      &$0.036$     &$0.016$             &$1.37$   \\
P0514361025 &59965.369 &$1.26\pm0.02$ &$5.528$   &$3.353 \times 10^{-8}$     &$0.11$                                 &$2085$            &$12.3$    &$0.037$      &$0.007$     &$2\times10^{-5}$    &$0.43$   \\
P0514361035 &59973.111 &$2.27\pm0.02$ &$5.417$   &$0.895 \times 10^{-8}$     &$0.046$                                &$4119$            &$16.23$   &$0.046$      &$0.009$     &$1\times10^{-5}$    &$0.43$   \\
P0514361038 &59977.804 &$2.22\pm0.01$ &$5.597$   &$0.989 \times 10^{-8}$     &$0.073$                                &$3791$            &$15.50$   &$0.045$      &$0.008$     &$1\times10^{-5}$    &$0.42$   \\
P0514361045 &59987.917 &$2.40\pm0.22$ &$1.833$   &$0.012 \times 10^{-5}$     &$0.068$                                &$3304$            &$14.86$   &$0.043$      &$0.024$     &$0.12\times10^{-3}$ &$0.74$   \\
P0514361049 &59992.480 &$2.50\pm0.14$ &$2.080$   &$0.146 \times 10^{-5}$     &$1.03$                                 &$166$             &$5.58$    &$0.027$      &$0.013$     &$0.37\times10^{-3}$ &$0.69$   \\
P0514361057 &60005.769 &$1.62\pm0.08$ &$3.850$   &$0.031 \times 10^{-2}$     &$468.15$                               &$0.46$            &$0.75$    &$0.009$      &$0.002$     &$2.94\times10^{-3}$ &$0.51$   \\
P0514361067 &60021.773 &$0.50\pm0.01$ &$3.072$   &$0.267 \times 10^{-2}$     &$2702.5$                               &$0.19$            &$0.53$    &$0.007$      &$0.002$     &$0.011$             &$0.57$   \\
P0514361074 &60029.248 &$0.36\pm0.02$ &$0.590$   &$3.398 \times 10^{-2}$     &$471.85$                               &$0.34$            &$0.65$    &$0.008$      &$0.014$     &$0.200$             &$1.30$   \\
\hline
\end{tabular}
\tablefoot{The source luminosity $L_{x}$ is estimated in 1-120\kev, in units of $10^{37}$\lumcgs. 
The distance to the source is assumed to be 2.44\kpc \citep{2021AJ....161..147B}. The accretion mass flux, 
$J$ is in units of \gpcmsqps.}
\vspace{-10pt}
\end{table*}

The phase-averaged spectral studies from the outburst observation of the pulsar shows that, the low-energy spectra is formed with the additional contribution from black-body like emission from the low-energy X-ray photons. The temperature of soft blackbody component varied in the energy 0.24--0.55\kev across the giant outburst. Whereas the high temperature blackbody component, that is mostly seen during the peak and declining phases of the giant outburst, varies within $1.9-2.7$\kev. In terms of emitting region radius, it can be estimated from the expression relating to blackbody normalization as $N_{BB}=R_{km}^{2}/D_{10}^2$, where the source distance, $D_{10}$ is in units of 10\kpc and $R_{km}$ is the radius of the blackbody emitting region in km. It is estimated that, the soft blackbody has an emission size radius extending to $5-176$~km. The high temperature blackbody component has an emission radius varying within ($\sim$0.5--$\sim$1.0)km . \citet{2024A&A...689A.316L} have also found that, during the bright state, the emission from the cold blackbody component extends to 10--19 km in radii to which they relate its emission originating from the optically thick accretion curtain. However the origin of high temperature blackbody component which is atypical for BeXB pulsars remains speculative. The emission from hot blackbody component is also seen during the \astrosat\ observations of \rxj during the 2022-23 outburst at temperatures of $\sim1.6$\kev \citep{2024MNRAS.534.1028S}, which has been associated with its emission originating from the top of the accretion column as its emission radius extends to 1-2 km.  Apart from these two blackbody emission components in the bright X-ray spectra of \rxj, we have also detected two cyclotron resonance scattering features seen in hard X-rays varying in energies 33.6--41.6\kev \& 64.6--75.3 \kev (see section~\ref{avg-spec}). These features are seen typically in the hard X-ray spectra of more than a dozen of bright X-ray pulsars at energies $\simeq10$\kev to $\sim100$\kev \citep{2019A&A...622A..61S,2002ApJ...580..394C}. An estimate of the NS magnetic field can be obtained from the measured value of cyclotron line energy through the relation: \\
 \vspace{-10pt}
 \begin{equation}
     B_{NS}(10^{12}G)=\frac{1}{\sqrt{\left(1-\frac{2GM_{NS}}{R_{NS}c^2}\right)}}\frac{E_{cycl}(keV)}{11.6}.
 \end{equation}
 \vspace{-1pt}
We found it to be varying in the range $\simeq3.4\times10^{12} G$ to $\simeq4.2\times10^{12} G$ as obtained from the variable cyclotron line energy, where we have assumed the canonical values of NS mass, $M_{NS}=1.4$\msol and radius, $R_{NS}=10$~km. 
Based on the theoretical estimates, the expression for critical luminosity in terms of NS surface CRSF energy as per \citet{2012A&A...544A.123B} is
\(L_{crit}=1.28\times10^{37}erg.s^{-1}\left(\frac{E_{cycl}}{10~keV}\right)^{16/15}\), where $E_{cycl}$ is the CRSF energy in \kev. Considering the variable CRSF energy as described above, we estimate the critical luminosity regime in the range of : $4.67$\lumcgsts -- $5.85$\lumcgsts. We find that this value is comparatively higher than the critical luminosity estimate by \citep{2023MNRAS.524.5213S} based  spectral turnover in hardness-luminosity variations of \rxj during the present giant outburst.

Although, the empirical spectral continuum models could successfully describe the spectra across the wide luminosity range exhibited by pulsar accretion column emission, they fail to elucidate a complete physical scenario of the emission spectrum. The application of thermal and bulk Comptonization model \citep{2007ApJ...654..435B} to the phase-averaged and phase resolved spectra of \rxj has revealed some of the important characteristics of the pulsar accretion column emission. During the outburst peak episodes, we observe that, the column radiation is mostly thermally dominated. This is seen from the lower values of $\delta$ during most of the outburst (as shown in Fig.\ref{lsv-avg-bw}). Alternatively, the higher values of photon diffusion parameter, $\xi$ noticed before and after the peak outburst episode also hints at an elongated accretion column emission structure. From the application of BW model to the phase-averaged outburst spectra of \rxj, we find the evidence of strong magnetic field of the pulsar. BW model predict a high magnetic field strength of \ie $\sim1.54\times10^{13}G$  in the BeXB pulsar \rxj during the declining phases of the outburst.  We note that, from the observed accretion induced spin-up of the pulsar during the present outburst, \citet{2023MNRAS.524.5213S} also estimate a high magnetic field  value of the pulsar of $B\approx3.5\times10^{13}$~G. 

Since the pulsar in \rxj, reached the super-critical accretion regime during its recent outburst in 2022-2023, \citep{2024A&A...681A..25M} have report the appearance of a transient spin-phase dependent QPO at frequencies $\sim0.2$Hz from \fermi-GBM observations at source luminosity of 2.8\lumcgsts. As per Keplerian frequency model, it is suggested that, the detection of $\sim200$ mHz QPO during the supercritical regime of accretion imply a high magnetic field of the pulsar of $2.3\times10^{13}$G \citep{2024A&A...681A..25M}. In addition to this, \citet{2024MNRAS.529.1187L} found that, the QPO frequency seen during the flaring episodes at the pulse-profiles peak varies in the range of 200 -- 500 mHz and assumed its probable origin from instability seen in the accretion flow.  From these variations, they estimate a strong magnetic field of the pulsar with strength $B=(3.3\pm0.2)\times10^{13}$G. However, considering a lower limit on the break frequency, the minimum magnetic field strength of the pulsar is reported as $(5.0\pm0.3)\times10^{12}$G \citep{2024MNRAS.529.1187L}. From the observed QPOs  seen at 60 mHz and 42 mHz during the \astrosat\ observations of the pulsar, when the pulsar is accreting at its sub-critical luminosities of 2.6\lumcgsts and 1.5\lumcgsts respectively, \citep{2024MNRAS.534.1028S} also report the higher magnetic field of the pulsar to be $7.3\times10^{13}$ G and $8.4\times10^{13}$G. From the above discussion it can be seen that, the pulsar magnetic field estimated using different methods such as from the variations in mHz QPOs frequecy, from the rapid spin-up of pulsar during the giant outburst and from the study of phase-averaged and phase-resolved outburst spectra  with thermal and bulk Comptonization model imply a strong magnetic field of the pulsar in $\sim10^{13}$G. From the detection of variable CRSF features seen in the \hxmt observed outburst spectra of \rxj in the present work and  previously during the 2010 outburst \citep{2012MNRAS.421.2407T} and from setting a lower limit on the break frequency \citep{2024MNRAS.529.1187L}, the minimum magnetic field strength of the pulsar is estimated to be around $\sim10^{12}$G. These results suggest that there is a large spatial offset between the Comptonized emission region and the cyclotron absorption region within the accretion column. Such offsets are also observed in other highly magnetized pulsars such as \fuo that shown multiple cyclotron lines in its spectrum \citep{2009A&A...498..825F}. 

Based on the application of Comptonization model \citep{2007ApJ...654..435B} to the phase-averaged outburst spectra of \rxj observed at different source luminosity, we could compute some of the additional physical parameters associated with BW model. In Table~\ref{t7}, we estimate some these parameter variations seen at different instances of the giant outburst. The parameters are : (1) $\alpha$ , a positive constant for the assumed velocity profile of the form $v(\tau)=-\alpha c\tau$. It is of the order of unity for the radiation dominated pulsar accretion column \citep{1998ApJ...498..790B,2007ApJ...654..435B}, (2)$\sigpar/\sig$: scattering cross sections for photons propagating parallel to the magnetic field in terms of Thomson scattering cross-section ($\sig$),  (3) $\sigbar/\sig$: Angle averaged scattering cross section in terms of $\sig$, (4) $J$: mass accretion flux (in units of \gpcmsqps), (5) $\Tmound$($\times10^{7}$K): thermal mound temperature (in units of $10^{7}K$, (6) $\vmound/c$: the inflow speed at the mound surface in terms of $c$, (7) $\taumound$: optical depth at the top of the thermal mound, (8) $\taumax$: the optical depth at the top of the radiating zone within the accretion column and (9) $\tautrap$: optical depth for the photons for the radiation to be trapped in the lower regions of accretion column, where generally $\tau<\tautrap$. And the radiation is transported vertically in the column. These parameters are defined respectively in the equations (33), (79), (83), (84), (88), (89), (92), (93) and (107) of \citet{2007ApJ...654..435B}. 

As shown in Table~\ref{t7}, the inflow speed of the accreting matter reaching the NS thermal mound surface, $\vmound$ during the onset and final phases of the giant outburst is $<0.02c$. Whereas the free-fall velocity at the top of the accretion column is $v_{ff}\sim0.6c$. This represents a significant deceleration of accreting plasma at the thermal mound. During the peak outburst phases (\ie between MJD 59973 -- 59987) we see a relatively high value of $\vmound$ (\ie $\geq0.04c$) compared to initial and final stages of the outburst in \rxj. This could be due to formation of radiation dominated shock for the source luminosity reaching $L_{x}\sim10^{37-38}$\lumcgs, wherein the accreting matter is halted at a certain height above the NS surface, before it is settled down onto the NS surface. For a comparison, during the 2021 giant outburst of \exo at the peak luminosity of $9.65$\lumcgsts, the inflow speed at the NS thermal mound was $<0.02c$\citep{2024ApJ...969..107Y}. The accretion velocity reduces by a factor of $\sim7$  while the gas passes through the shock \citep{2012A&A...544A.123B}. At the peak luminosity, displayed by the pulsar in \rxj during the giant outburst, the radiation field could decelerate the accreting gas all the way down to the NS surface. 

The temperature of the gas in the thermal mound ($T_{th}$) increases with the progress of the outburst, as the source luminosity and mass accretion flux ($J$) increase. At the peak episodes, $T_{th}$ varies at higher values of  (14 -- 16)$\times10^{7}K$. As expected the electron temperature at the top of the thermal mound, $T_e$ at these episodes is also found to be of similar values and comparable to $T_{th}$ (as shown in Figure~\ref{lsv-avg-bw}).
From the comparison of $\taumax$ and $\tautrap$ values obtained at different luminosity states of the pulsar \rxj during the giant outburst, we find that the condition $\taumax\leq\tautrap$ is valid. This shows that, the ``trapped'' region of accretion column is from where the most of the observed emission from the pulsar is produced during the giant outburst. In Table~\ref{t7}, we have also estimated the variations seen in two scattering cross section values (\ie $\sigpar/\sig$, $\sigbar/\sig$) with the outburst source luminosity. The $\sigpar/\sig$ denote the scattering cross section for the photon propagating parallel to the magnetic field in-terms of Thomson scattering ($\sig$), whereas the $\sigbar/\sig$ is angle-averaged cross sections in units of $\sig$. The mean scattering cross section for photons propagating perpendicular to the field is usually set to $\sig$, \ie $\sigperp\approx\sig$ \citep{2007ApJ...654..435B}. At the peak luminosities, we find that $\sigpar<<\sigbar<<\sigperp$. \ie The radiation is subjected to a reduced scattering cross-sections, which prefers the photons to travel along the magnetic field lines. Thus the accreting material in the column could decelerate to rest at the NS surface instead of being blown away \citep{1979PhRvD..19.1684V,1971PhRvD...3.2303C,2007ApJ...654..435B,2009A&A...498..825F}. The smaller values of column emission radius $r_0<50 m$ during the outburst peak (as shown in Figure~\ref{lsv-avg-bw}), also indicates that, the photons almost travel to the base of the accretion column before escaping from the column walls where the plasma temperature is higher.

\section{Conclusions}
In this paper, we have analyzed all the archival \hxmt\ observations of the BeXB pulsar \rxj during its recent giant outburst in late 2022-2023. The pulse profiles during the progress and declining phases of the outburst at similar luminosities show similar variations in the shape, which indicates a strong luminosity dependence of pulse profiles in \rxj. The net X-ray luminosity of the pulsar during the giant outburst spanned in the range  $4.4\times10^{36}$\lumcgs -- $2.8$\lumcgsts in 1--120\kev assuming a distance of 2.44\kpc. The phase-averaged and phase-resolved spectra of the pulsar at different luminosity states could be explained with an absorbed \cpl\ continuum model along with two additional blackbody components at soft X-rays. The high cadence hard X-ray observations of the pulsar reveal the presence of a variable CRSF in the energies of (33.6--41.6)\kev. During the declining phases of the giant outburst, we have also discovered a second cyclotron line varying in the range of $\sim64.6$ -- 75.3\kev. The presence of variable cold blackbody component in ($0.24-0.55$)\kev seen in the phase-averaged, and phase-resolved spectral studies implies its origin to be from the optically thick accretion curtain as the size of its emission region extends to $(5-176)$km. Whereas the variation of a hot blackbody component seen in ($1.9-2.7$)\kev\, which is atypical of BeXB pulsars, has an emission size within ($0.5-1$)km originating from the upper regions of accretion column. We have also used the physics-based thermal and bulk Comptonization model to explore the accretion column emission at many luminosity epochs of the pulsar during the giant outburst. These studies reveal a strong magnetic field of the pulsar of the order of $\sim10^{13}$G observed during peak luminosities. Most of the Comptonized emission originated from the narrow and lower regions of NS accretion column emission since the column emission radius at outburst peak episodes is found to be $< 50$ m . 

\begin{acknowledgements}
We are grateful to the referee for the comments. This work is supported by the NSFC (No. 12133007) and National Key Research and Development Program of China (Grants No. 2021YFA0718503). 
\end{acknowledgements}

\bibliographystyle{aa}
\bibliography{bibtex/aa54715-25}

\begin{appendix}
\onecolumn
\section{Observation log}
{\small
 \begin{longtable}{cccccccc}
 \caption{Observation IDs of the \hxmt observations covering 2022-2023 Giant outburst of \rxj.}
 \label{tab:log1}\\
 \hline
 \multirow{2}{*}{ObsID} &\multirow{2}{*}{Time Start (UTC)}  &\multirow{2}{*}{\thead{Exposure time (s) \\ LE/ME/HE}} &\multirow{2}{*}{MJD} &\multirow{2}{*}{\thead{Flux(1-120~\kev)$^a$ \\ (10$^{-9}$ erg cm$^{-2}$ s$^{-1}$)}}  &\multirow{2}{*}{\thead{PF$^b$(\%) \\ (1-10 \kev)}} &\multirow{2}{*}{\thead{PF$^b$(\%) \\ (10-30 \kev)}} &\multirow{2}{*}{\thead{PF$^b$(\%) \\ (30-150 \kev)}} \\\\\\ 
 \hline
 \endfirsthead
 \caption{continued.}\\
 \hline
 \multirow{2}{*}{ObsID} & \multirow{2}{*}{Time Start (UTC)}  & \multirow{2}{*}{\thead{Exposure time (s) \\ LE/ME/HE}} & \multirow{2}{*}{MJD} & \multirow{2}{*}{\thead{Flux(1-120\kev)$^a$ \\ (10$^{-9}$ erg cm$^{-2}$ s$^{-1}$)}}  & \multirow{2}{*}{\thead{PF$^b$(\%) \\ (1-10 \kev)}} & \multirow{2}{*}{\thead{PF (\%) \\ (10-30\kev)}} & \multirow{2}{*}{\thead{PF(\%) \\ (30-150 \kev)}} \\
 \hline
 \endhead
 \hline
 \endfoot
P0514361001  &2022-12-31   &2556/13290/11950  &59944.40  &$18.55^{+1.58}_{-1.42}$  &$44.12\pm0.90$  &$30.40\pm0.38$   &$38.70\pm0.42$ \\ 
P0514361002  &2023-01-01   &1676/8868/7128    &59945.46  &$18.86^{+1.85}_{-1.67}$  &$43.90\pm1.00$  &$32.83\pm0.43$   &$38.98\pm0.53$ \\  
P0514361003  &2023-01-02   &1601/7872/6919    &59946.65  &$16.28^{+1.62}_{-1.46}$  &$44.66\pm0.95$  &$32.38\pm0.46$   &$31.10\pm0.56$ \\  
P0514361004  &2023-01-03   &1736/9217/6038    &59947.44  &$17.24^{+1.75}_{-1.57}$  &$43.82\pm0.96$  &$31.90\pm0.43$   &$35.71\pm0.55$ \\ 
P0514361005  &2023-01-04   &1469/7370/7313    &59948.63  &$16.56^{+1.82}_{-1.42}$  &$46.22\pm1.03$  &$31.00\pm0.50$   &$36.04\pm0.73$ \\  
P0514361015  &2023-01-09   &598.5/6169/6021   &59953.13  &$10.76^{+3.01}_{-2.08}$  &$45.80\pm2.19$  &$34.17\pm0.81$   &$45.09\pm1.64$ \\
P0514361016  &2023-01-10   &419/6068/7578     &59954.72  &$7.03^{+2.33}_{-1.58}$  &$48.43\pm4.82$  &$32.24\pm0.93$   &$44.08\pm1.61$ \\ 
P0514361017  &2023-01-11   &1059/5717/4680    &59955.51  &$7.11^{+1.87}_{-1.42}$  &$46.57\pm2.53$  &$32.37\pm1.34$   &$36.08\pm2.85$ \\ 
P0514361018  &2023-01-12   &830.9/4553/6059   &59956.70  &$6.25^{+2.00}_{-1.48}$  &$40.05\pm3.01$  &$32.67\pm1.39$   &$49.41\pm3.08$ \\  
P0514361019  &2023-01-13   &2585/5178/7551    &59957.63  &$6.39^{+1.40}_{-1.13}$  &$50.95\pm2.03$  &$35.94\pm1.53$   &$55.65\pm3.63$ \\  
P0514361022  &2023-01-17   &1166/3365/3330    &59961.86  &$30.06^{+5.89}_{-4.84}$  &$41.84\pm1.13$  &$32.21\pm0.67$   &$46.81\pm0.96$ \\
P0514361023  &2023-01-18   &3612/4514/5922    &59962.65  &$15.19^{+3.17}_{-2.92}$  &$44.16\pm0.65$  &$34.34\pm0.51$   &$35.43\pm0.53$ \\ 
P0514361024  &2023-01-20   &3520/6991/5861    &59964.04  &$19.20^{+2.12}_{-1.91}$  &$51.20\pm0.52$  &$36.41\pm0.36$   &$35.16\pm0.42$  \\ 
P0514361025  &2023-01-21   &4535/5623/5125    &59965.37  &$17.09^{+1.24}_{-0.90}$  &$47.62\pm0.64$  &$33.97\pm0.37$   &$39.20\pm0.40$ \\ 
P0514361026  &2023-01-22   &3369/3612/5311    &59966.50  &$21.62^{+2.91}_{-3.42}$  &$50.33\pm0.49$  &$36.54\pm0.44$   &$39.04\pm0.36$ \\ 
P0514361027  &2023-01-23   &2918/3141/4266    &59967.56  &$26.32^{+3.14}_{-2.82}$  &$54.38\pm0.54$  &$39.86\pm0.46$   &$42.70\pm0.37$ \\ 
P0514361028  &2023-01-24   &5222/5719/6215    &59968.35  &$21.34^{+1.83}_{-1.71}$  &$51.42\pm0.42$  &$39.99\pm0.31$   &$43.55\pm0.30$ \\ 
P0514361029  &2023-01-25   &9677/11670/11770  &59969.01  &$23.13^{+1.35}_{-1.30}$  &$55.85\pm0.28$  &$39.79\pm0.21$   &$45.77\pm0.22$ \\ 
P0514361030  &2023-01-26   &8438/11330/12040  &59970.00  &$23.85^{+2.81}_{-3.45}$  &$58.13\pm0.28$  &$40.40\pm0.21$   &$49.44\pm0.18$ \\  
P0514361031  &2023-01-27   &5995/6957/6806    &59971.26  &$27.28^{+2.15}_{-1.97}$  &$55.25\pm0.32$  &$42.34\pm0.24$   &$52.69\pm0.20$ \\ 
P0514361032  &2023-01-28   &5566/6745/7951    &59972.32  &$31.00^{+1.08}_{-1.00}$  &$57.43\pm0.28$  &$47.65\pm0.22$   &$56.78\pm0.18$ \\ 
P0514361035	 &2023-01-29   &22200/35510/35410 &59973.11  &$33.74^{+0.77}_{-0.72}$  &$51.77\pm0.14$  &$42.76\pm0.09$   &$52.42\pm0.09$ \\
P0514361036	 &2023-01-31   &7176/15690/16050  &59975.49	 &$32.16^{+0.24}_{-0.24}$  &$57.28\pm0.23$  &$49.67\pm0.13$   &$58.14\pm0.12$ \\ 
P0514361037	 &2023-02-01   &11230/20680/18800 &59976.48  &$32.10^{+0.24}_{-0.24}$  &$61.04\pm0.17$  &$52.83\pm0.11$   &$65.19\pm0.10$ \\
P0514361038	 &2023-02-02   &29970/61060/58130 &59977.80  &$31.05^{+0.32}_{-0.31}$  &$53.22\pm0.22$  &$53.77\pm0.27$   &$57.12\pm0.30$ \\
P0514361040	 &2023-02-05   &6900/13560/10830  &59980.71  &$30.16^{+0.62}_{-0.61}$  &$64.04\pm0.22$  &$58.73\pm0.13$   &$69.38\pm0.13$ \\
P0514361041	 &2023-02-06   &5765/8863/8135	  &59981.77	 &$29.42^{+0.73}_{-0.73}$  &$64.14\pm0.25$  &$57.82\pm0.16$   &$70.72\pm0.15$ \\
P0514361042	 &2023-02-07   &5419/7325/7207 	  &59982.76  &$29.23^{+1.70}_{-1.49}$  &$64.18\pm0.28$  &$58.38\pm0.18$   &$71.44\pm0.16$ \\
P0514361043	 &2023-02-08   &12500/23260/24940 &59983.95  &$29.48^{+1.43}_{-1.32}$  &$62.25\pm0.19$  &$54.98\pm0.11$   &$66.70\pm0.09$ \\
P0514361044	 &2023-02-10   &8861/15880/15410  &59985.87  &$31.05^{+1.37}_{-1.27}$  &$61.97\pm0.23$  &$53.74\pm0.14$   &$67.33\pm0.12$ \\
P0514361045	 &2023-02-12   &1077/3717/3559	  &59987.92  &$31.99^{+2.74}_{-2.45}$  &$61.86\pm0.67$  &$53.53\pm0.30$   &$65.63\pm0.25$ \\
P0514361046	 &2023-02-13   &1038/4218/2910	  &59988.91  &$27.78^{+1.22}_{-1.16}$  &$60.78\pm0.64$  &$53.29\pm0.28$   &$66.22\pm0.28$ \\
P0514361047	 &2023-02-14   &1506/4426/4842	  &59989.90  &$27.68^{+1.35}_{-1.30}$  &$57.45\pm0.64$  &$49.68\pm0.29$   &$61.96\pm0.23$ \\
P0514361048	 &2023-02-15   &2795/6348/6389	  &59990.83  &$37.57^{+3.44}_{-3.46}$  &$57.10\pm0.47$  &$47.35\pm0.24$   &$57.03\pm0.20$ \\
P0514361049	 &2023-02-17   &1196/5263/3478	  &59992.48	 &$34.81^{+15.96}_{-6.41}$	&$54.53\pm0.77$  &$43.52\pm0.29$   &$48.45\pm0.34$ \\
P0514361050	 &2023-02-18   &1831/5465/4994	  &59993.80  &$27.75^{+1.14}_{-1.09}$	&$54.01\pm0.57$  &$44.87\pm0.28$   &$50.42\pm0.26$ \\
P0514361051	 &2023-02-19   &2040/4091/4266	  &59994.86  &$39.31^{+2.72}_{-2.51}$	&$55.55\pm0.54$  &$44.91\pm0.33$   &$48.95\pm0.30$ \\
P0514361052	 &2023-02-20   &940.6/5663/7660	  &59995.92  &$28.65^{+1.92}_{-1.77}$	&$58.92\pm0.97$  &$48.98\pm0.31$   &$47.07\pm0.28$ \\
P0514361053	 &2023-02-23   &3052/18500/21180  &59998.04  &$34.84^{+1.65}_{-1.54}$	&$53.70\pm0.44$  &$46.28\pm0.16$   &$47.88\pm0.14$ \\
P0514361054	 &2023-02-24   &3405/17200/17370  &59999.49  &$36.34^{+1.83}_{-1.71}$	&$51.39\pm0.39$  &$44.28\pm0.16$   &$44.42\pm0.15$ \\
P0514361055	 &2023-02-26   &7333/26460/18190  &60001.14  &$22.18^{+1.99}_{-1.99}$	&$55.40\pm0.29$  &$43.27\pm0.14$   &$41.11\pm0.17$ \\
P0514361056	 &2023-02-28   &6291/20130/21640  &60003.98  &$18.59^{+0.62}_{-0.62}$	&$52.40\pm0.36$  &$40.90\pm0.18$   &$35.08\pm0.18$ \\
P0514361057	 &2023-03-02   &3920/7867/10620	  &60005.77  &$17.34^{+0.72}_{-0.70}$	&$52.94\pm0.46$  &$42.89\pm0.29$   &$33.53\pm0.28$ \\
P0514361058	 &2023-03-03   &3365/7134/9098	  &60006.69  &$16.08^{+1.24}_{-1.40}$	&$47.21\pm0.53$  &$35.38\pm0.33$   &$34.57\pm0.32$ \\
P0514361059	 &2023-03-04   &2069/6181/9299	  &60007.69  &$15.68^{+0.93}_{-0.90}$	&$50.36\pm0.66$  &$37.77\pm0.36$   &$30.42\pm0.34$ \\
P0514361060	 &2023-03-05   &4666/8011/8855	  &60008.61  &$14.64^{+0.63}_{-0.62}$	&$51.35\pm0.44$  &$35.73\pm0.33$   &$30.42\pm0.36$ \\
P0514361061	 &2023-03-06   &2453/6756/8197	  &60009.87  &$13.41^{+1.28}_{-1.35}$	&$50.03\pm0.70$  &$34.15\pm0.38$   &$31.15\pm0.43$ \\
P0514361062	 &2023-03-07   &3062/8401/9526	  &60010.79  &$12.87^{+0.97}_{-0.98}$	&$48.82\pm0.64$  &$31.99\pm0.35$   &$27.67\pm0.39$ \\
P0514361064	 &2023-03-13   &1812/3832/4191	  &60016.81  &$9.68^{+0.78}_{-0.75}$	&$44.61\pm0.97$  &$30.25\pm0.64$   &$39.88\pm0.81$ \\
P0514361065	 &2023-03-15   &3352/9127/6893	  &60018.13  &$9.02^{+0.55}_{-0.53}$	&$39.47\pm0.75$  &$25.32\pm0.46$   &$40.03\pm0.73$ \\
P0514361066	 &2023-03-16   &3292/8080/7278	  &60019.99  &$7.97^{+0.51}_{-0.49}$	&$46.70\pm0.79$  &$26.43\pm0.52$   &$37.28\pm0.81$ \\
P0514361067	 &2023-03-18   &1137/4822/4367	  &60021.77  &$7.44^{+0.65}_{-0.61}$	&$36.66\pm1.75$  &$22.97\pm0.78$   &$42.07\pm1.23$ \\
P0514361068	 &2023-03-19   &1809/6872/6719	  &60022.96  &$10.45^{+3.08}_{-2.71}$	&$34.04\pm1.17$  &$23.60\pm0.63$   &$41.19\pm1.05$ \\
P0514361071	 &2023-03-21   &2609/3760/4936	  &60024.55  &$9.12^{+2.31}_{-2.22}$	&$33.33\pm1.02$  &$22.05\pm0.83$   &$40.46\pm1.13$ \\   
P0514361072	 &2023-03-23   &2977/8043/6296	  &60026.01  &$12.59^{+2.81}_{-2.29}$	&$30.13\pm0.98$  &$19.39\pm0.63$   &$38.44\pm1.27$ \\
P0514361073	 &2023-03-24   &1542/4885/5402	  &60027.79  &$11.75^{+3.35}_{-2.43}$	&$32.31\pm1.34$  &$19.43\pm0.84$   &$40.79\pm1.35$ \\
P0514361074	 &2023-03-26   &5105/4948/5490	  &60029.25  &$9.98^{+1.68}_{-1.40}$	&$26.07\pm0.81$  &$20.45\pm0.87$   &$41.59\pm1.50$ \\
\end{longtable}
\begin{tablenotes}
\item $^a$ The unabsorbed flux of \rxj is computed from the best-fit \cpl\ model in the energy range 1-120 keV (see Section \cref{avg-spec}). 
\item $^b$ The pulse fraction (PF=($I_{max}$-$I_{min}$)/($I_{max}$+$I_{min}$)) is calculated from the background-subtracted light curves. The error on the pulse fraction estimates is obtained from the propagation of errors.
\end{tablenotes}
}
\twocolumn
\onecolumn
\section{Additional figures}

\begin{figure*}[!ht]
 \begin{center}$
 \begin{array}{c}
 \includegraphics[height=3.4in, width=4.5in, angle=0]{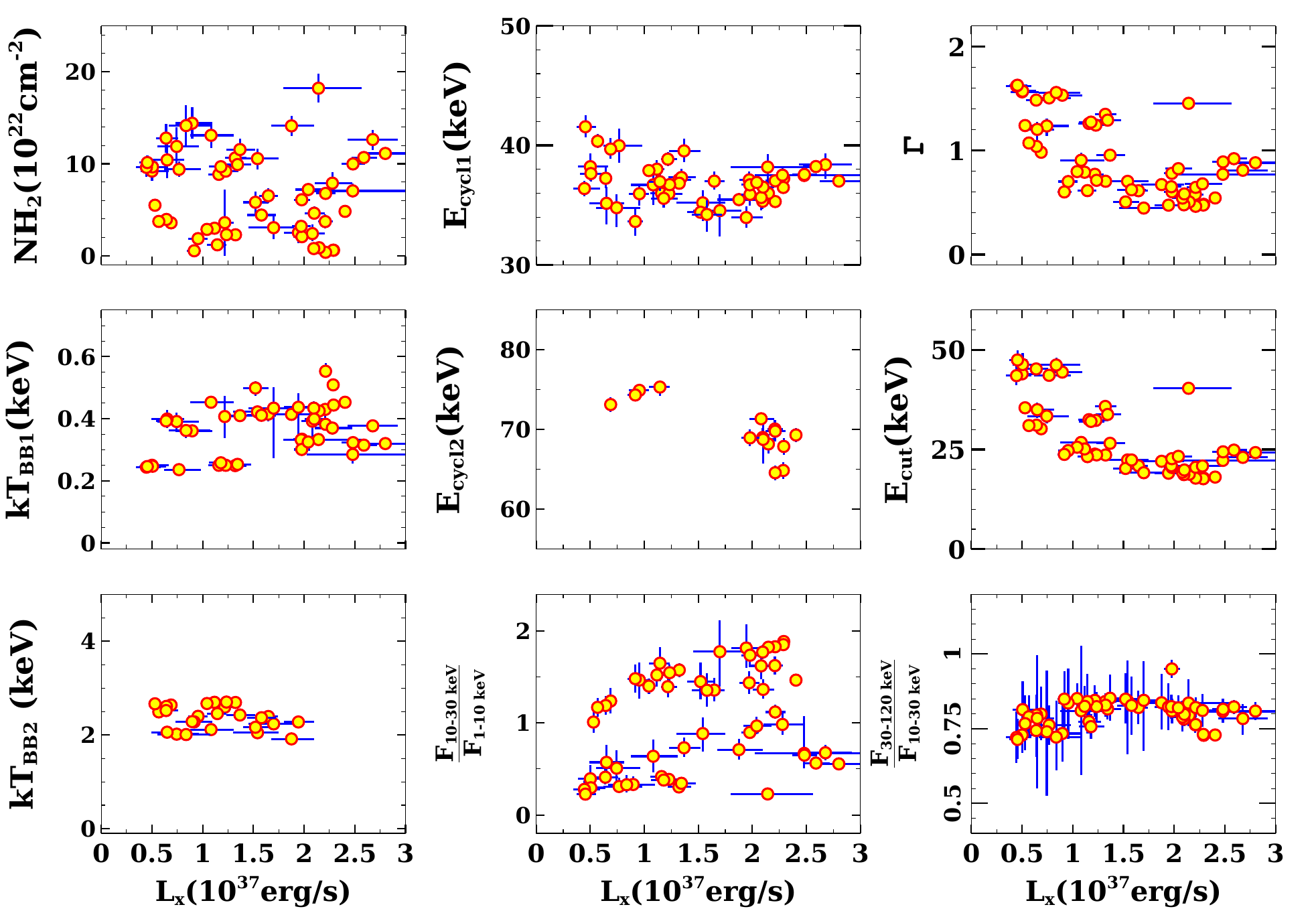} 
 \end{array}$
 \end{center}
\caption{\small Variation of spectral parameters of \rxj w.r.t outburst source luminosity in 1--120\kev of the giant outburst. An absorbed cutoff power-law model (\ie \cpl) with additional components like: 1) a Gaussian component at 6.4\kev, 2) a cyclotron absorption component at ~35-40\kev, 3) a soft black-body component near 0.3\kev, and sometimes a  second blackbody component near 2.5\kev has been used to obtain the best-fitted continuum model.}
\label{lsv-phase-avg-2}
\end{figure*}  

\begin{figure*}
 \begin{center}$
  \begin{array}{c}
   \includegraphics[height=3.4in, width=4.5in, angle=0]{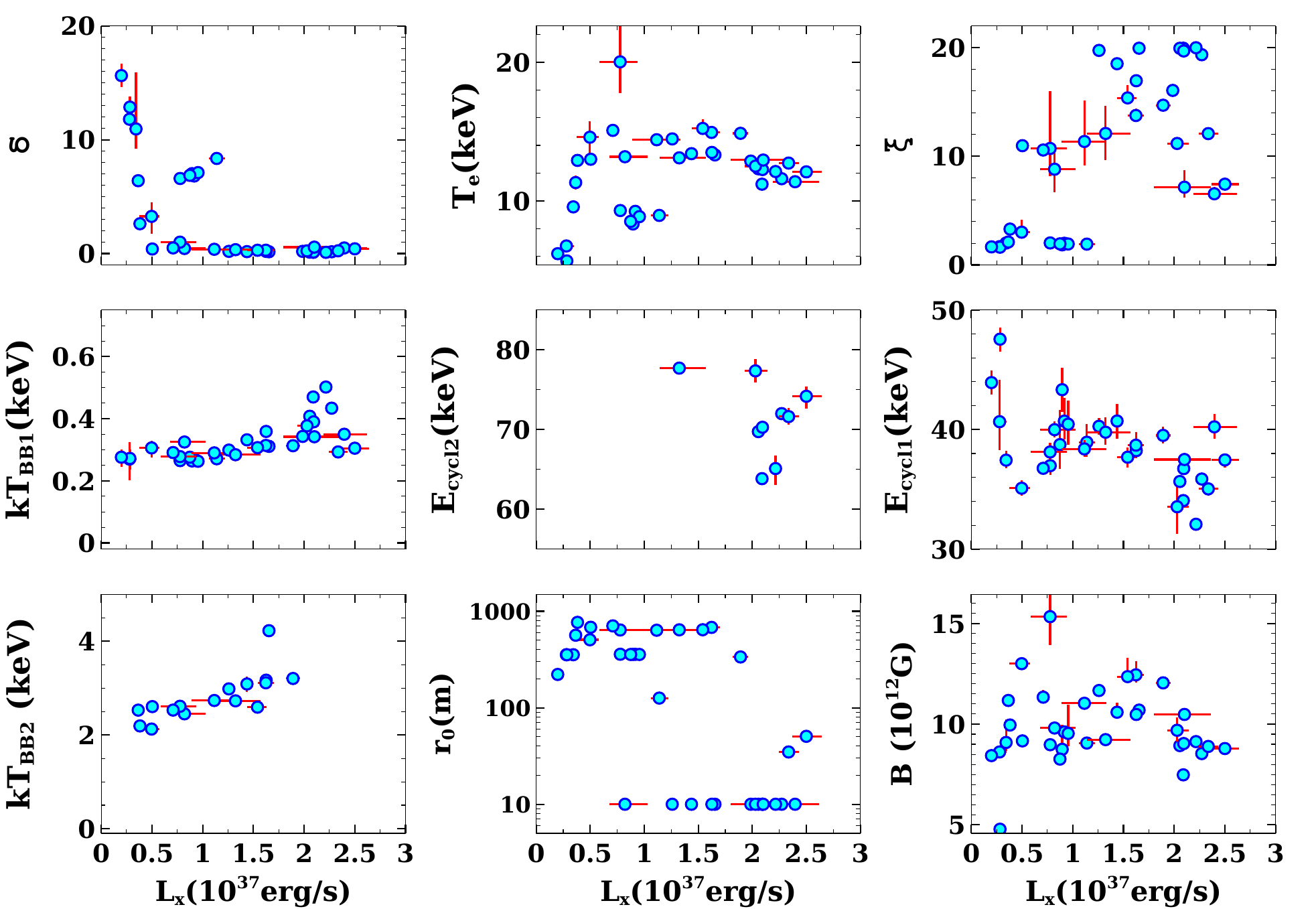}
 \end{array}$
 \end{center}
 \caption{\small The best-fitted phase averaged spectral parameters using \emph{bwcycl} model obtained at different epochs of the the giant outburst in \rxj as observed with \hxmt. The spectral variations is shown with variation in source luminosity across the giant outburst. The error-bars represent 1$\sigma$ uncertainties.}
 \label{lsv-avg-bw-2}
\end{figure*}  
\FloatBarrier
\twocolumn

\section{Phase-resolved spectroscopy with BW Model}\label{phase-res-BW}

\begin{figure*}[ht]
 \begin{center}
 \includegraphics[height=5.5in, width=7.0in, angle=0]{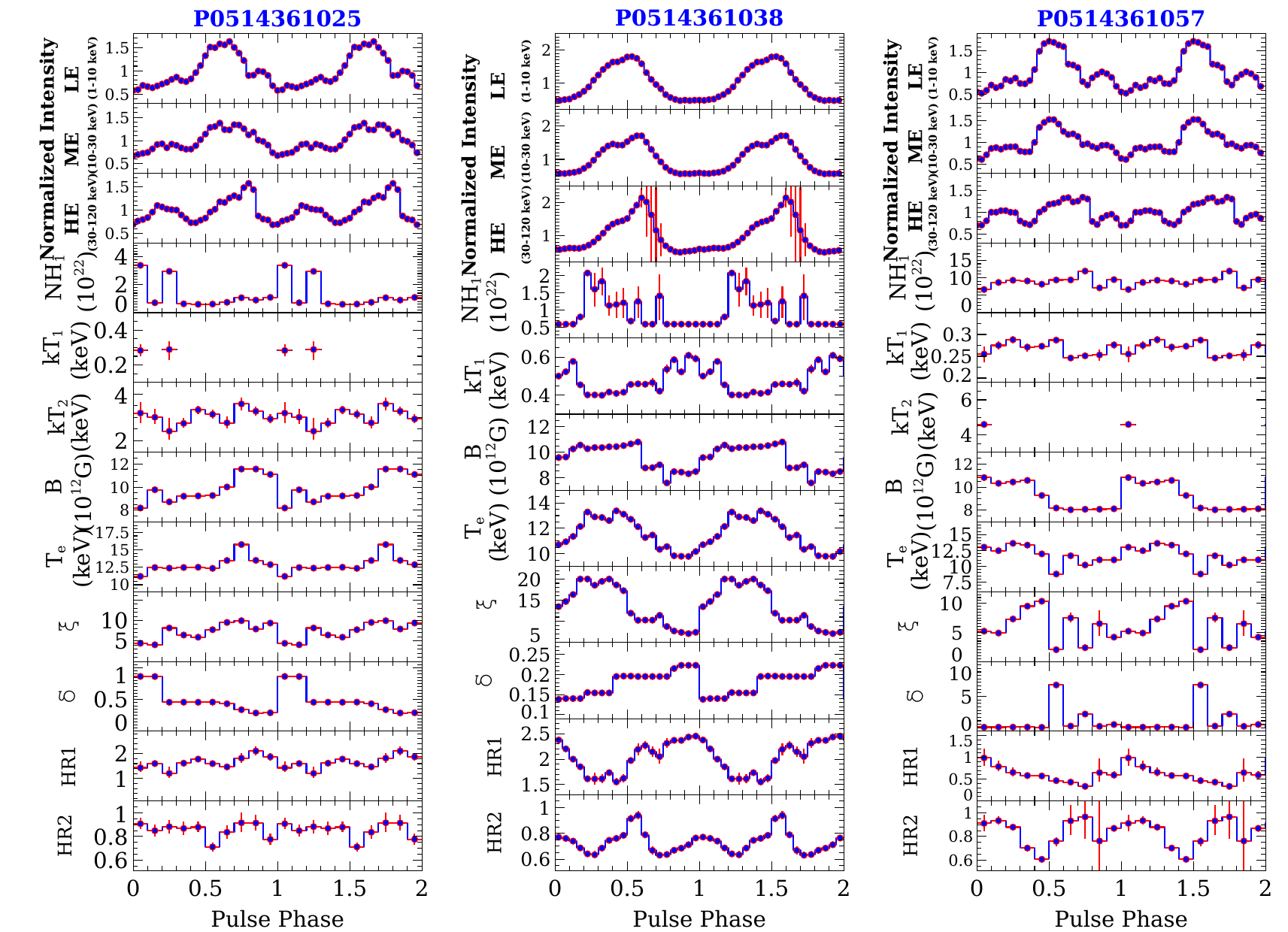} 
 \end{center}
\caption{\small Variation of spectral parameters with pulse-phases as obtained with a Comptonization model (\ie \emph{bwcycl}) at three different luminosity epochs. The error-bars represent 1$\sigma$ uncertainties. The observation IDs are shown on the top of each panel. Corresponding epochs of the \hxmt observations can be seen from Table~\ref{tab:log1} in the Appendix.}
\label{lsv-res-bw}
\end{figure*}

We also perform the phase-resolved spectral studies with the BW07 model at the three luminosity epochs as it has been done with the empirical \cpl\  model in section~\ref{sec:empirical-phase-res}. We have used the same three \hxmt observations, to carry out phase resolved spectroscopy with BW07 model, as those have been used during phase-resolved spectral studies with the \cpl\ model. These observations are namely with ObsdIDs: P0514361025 (during the progress of the outburst), P0514361038 (near the outburst peak) and P0514361057 (at the decline of the outburst, with the similar source luminosity as during the ObsID P0514361025). The broadband spectra of the pulsar obtained from LE, ME and HE during observations (\ie with ObsIDs: P0514361025, P0514361057) have been divided into 10 pulse-phase bins. Whereas the spectra obtained from ObsID: P0514361038, has been divided into 20 pulse phase bins. To obtain the  phase-resolved spectra, we have considered the pulsar spin-period obtained obtained near the epochs of each observation \citep{2023MNRAS.526.3637L} and folding epochs chosen near each observation such that two complete pulse profiles (in ME: 10--30\kev) can be seen in 0.0-2.0 pulse-phases.  In the top 3 panels of the Figure~\ref{lsv-res-bw}, we show the pulse-profiles obtained from the light curves in (1--10)\kev, (10--30)\kev and (30--150)\kev respectively for each of the observed epochs.  

The BW07 physical model considers a vertically integrated accretion column emission spectrum that is independent of the viewing angle of observer in the rest frame of the neutron star \citep{2009A&A...498..825F, 2007ApJ...654..435B}. However, its angular dependence emerges from the fact that, it considers the simplified approach of the scattering cross sections for photon propagating in parallel ($\sigpar$) and perpendicular ($\sigperp$) to the field.  Therefore a study of the phase-resolved spectrum of the pulsar could illustrate the variation of scattering cross sections with pulse phases through the variations in the BW physical parameter $\xi$ (as per equation~\ref{eq:eq1}). Also the dependency of other physical parameters such as $B$, $T_e$ \& $\delta$ with pulse-phases can be studied to understand the pulse-phase variations of column emission. 

While fitting with BW07 model to the phase-resolved spectra, we fix the values of \mdot and $r_0$ to the values that are obtained during the phase averaged spectral studies. This leaves us to check the variation of other physical BW07 model best-fit parameters  such as $B_{12}$, $T_e$, $\xi$ and $\delta$ with pulse phases of the pulsar. Apart from the BW07 model component, we also require two additional \bbody\ components with temperatures \kt$\sim0.3$\kev and \ktt$\sim3-5$\kev to better fit the spectra at certain pulse phases.  In hard X-rays, we do require the additional absorption components namely \cyclabs\ to account for the variations of CRSF features seen across pulse phases. The pulse-phase variations of CRSF are found similar to the results obtained during the phase-resolved spectral studies with the best-fit empirical \cpl\ model as described in section~\ref{sec:empirical-phase-res}. In Figure~\ref{lsv-res-bw}, we show the phase dependency of the best-fit BW07 model parameters across pulse phases of the pulsar in \rxj during the three epochs of the giant outburst observations with \hxmt.  We notice that, the magnetic field, $B_{12}$ is higher during the peak-intensity phases of the pulse-profile (\ie between 0.2--0.7 pulse phase) during the outburst peak observation (\ie ObsID: P0514361038), whereas during the other two epochs, it is found to be higher at the edges of the main peak of the pulse profile.  From the variation of $\delta$ parameter with pulse phases, we find the contribution of bulk Comptonization w.r.t thermal Comptonization is minimal during the outburst peak as $\delta$ was varying within 0.14--0.22. During the progress of the outburst (\ie in ObsID: P0514361025), we find that, $\delta$ was varying within 0.29--0.86, with the higher value observed during the 0.0--0.2 pulse-phases where the minor sub-peak in the pulse profile is seen. This indicates that the contribution of bulk Comptonization w.r.t thermal Comptonization is moderate during the progress of the giant outburst. However, during the declining phases of the outburst (\ie as in ObsID: P0514361057), we find the $\delta$ parameter to be highly variable, reaching higher values of $\sim6.63$ at peak-intensity phases (\ie 0.5--0.6 pulse phases) of the pulsar. This indicates that contribution from bulk Comptonization is very high at the main peak of the pulse profile. 

During the other pulse phases, we find the relative contribution from thermal Comptonization leading the energization of X-ray photons to higher energies. Moreover, the parameter $\xi$ follows the luminosity phase dependency during the giant outburst peak. In particular, during the outburst peak, we observe $\xi$ reaching values $>15$, within 0.1-0.5 pulse phases of the main peak. This implies that the trapping altitude ($z_{trap})$ for photon diffusing vertically in the column is more than 30 times the distance of the sonic point from the NS surface ($z_{sp}$). This makes it feasible for photons to reach higher heights of accretion column emission. However during the pulse phases 0.5-1.0, we see a substantial decrease in the $\xi$ values to 5. This implies that, the Comptonized emission from the pulsar is viewed from different altitudes in the accretion column across the pulse phases or the viewing angle of accretion column is highly variable across the pulse-phases. Similar variations we have seen in case of electron plasma temperature, $T_e$ with pulse phases.  It is higher during the main peak of the pulse profiles, reaching as high as $\sim13.4$\kev (during ObsID: P0514361038). However average plasma temperature during this outburst peak observation is ($12.2\pm0.2$)\kev, which is lower than the average $T_e$ values seen during the progress and decline phases of the outburst as observed during ObsID: P0514361025 ($T_e=14.5\pm0.2$)\kev and ObsID: P0514361057($T_e=14.9\pm0.4$)\kev respectively. This hints a partial escape of photons through column walls, leading to cooling of plasma in the accretion column at the outburst peak luminosity episodes. The variation of two flux ratios with pulse phases as shown in two lower panels in Figure~\ref{lsv-res-bw} for each of the observed epoch has shown similar variations across pulse phases, as it has been observed during the phase-resolved spectral studies with best fit \cpl\ model (see Section~\ref{sec:empirical-phase-res}). 
\FloatBarrier 
\twocolumn
\end{appendix}
\end{document}